\documentclass[11pt]{article}
\usepackage{epsf}

 at 10 truept
 at 10 truept
\font\appauthcs=cmcsc10 scaled 1095 
\font\titlefont=cmr7 scaled 1728

\makeatletter
\def\section{\@startsection{section}{1}{\z@}{3.5ex plus-1ex minus
    -.2ex}{0.1ex plus.1ex}{\reset@font\normalsize\bf}}
\def\subsection{\@startsection{subsection}{2}{\z@}{3.25ex plus-1ex
     minus-.2ex}{1ex plus.2ex}{\reset@font\normalsize\it}}
\def\subsubsection{\@startsection{subsubsection}{3}{\z@}{-3.25ex plus
     -1ex minus-.2ex}{1.5ex plus.2ex}{\reset@font\normalsize\bf}}
\def\paragraph{\@startsection
     {paragraph}{4}{\z@}{3.25ex plus1ex minus.2ex}{-1em}{\reset@font
     \normalsize\bf}}
\def\subparagraph{\@startsection
     {subparagraph}{4}{\parindent}{3.25ex plus1ex minus
     .2ex}{-1em}{\reset@font\normalsize\bf}}

\def\@sect#1#2#3#4#5#6[#7]#8{\ifnum #2>\c@secnumdepth
     \let\@svsec\@empty\else
     \refstepcounter{#1}\edef\@svsec{\csname the#1\endcsname.~}\fi
     \@tempskipa #5\relax
      \ifdim \@tempskipa>\z@
        \begingroup #6\relax
\begin{center}
          \@hangfrom{\hskip #3 \@svsec\relax}{\interlinepenalty \@M #8\par}%
\end{center}
        \endgroup
       \csname #1mark\endcsname{#7}\addcontentsline
         {toc}{#1}{\ifnum #2>\c@secnumdepth \else
                      \protect\numberline{\csname the#1\endcsname}\fi
                    #7}\else
        \def\@svsechd{#6\hskip #3\relax  
                   \@svsec #8\csname #1mark\endcsname
                      {#7}\addcontentsline
                           {toc}{#1}{\ifnum #2>\c@secnumdepth \else
                             \protect\numberline{\csname the#1\endcsname}\fi
                       #7}}\fi
     \@xsect{#5}}

\def\refname{REFERENCES}
\def\thebibliography#1{%
\vskip24pt
\centerline{\refname}%
\list
{\@biblabel{\arabic{enumiv}}}%
{
    \settowidth\labelwidth{\@biblabel{#1}}%
    \leftmargin\labelwidth
    \advance\leftmargin\labelsep
    \usecounter{enumiv}%
    \let\p@enumiv\@empty
    \def\theenumiv{\arabic{enumiv}}%
    \parsep\parskip
    \small
}
    \def\newblock{\hskip .11em plus.33em minus.07em}%
    \sloppy\clubpenalty4000\widowpenalty4000
    \sfcode`\.=1000\relax}

\def\citen#1{\let\@@cite\@cite%
\gdef\@cite##1##2{{##1\if@tempswa , ##2\fi}}%
\cite#1%
\let\@cite\@@cite}
\makeatother

\newcommand{\rr}[4]{{#1, }{\it #2\/ }{\bf #3}, \rm #4}
\newcommand{\rre}[4]{#1, #2 #4}

\begin{document}
\def\no{\nonumber}
\def\qu{\quad}
\def\qb{\bar{q}}
\def\qbm{\bar{\mbox{q}}}
\def\la{\langle}
\newcommand{\cor}[1]{\left<{#1}\right>}
\newcommand{\gm}{\gamma}
\newcommand{\be}{\begin{eqnarray}}
\newcommand{\ee}{\end{eqnarray}}
\renewcommand{\th}{\theta}
\newcommand{\Sg}{\Sigma}
\newcommand{\dl}{\delta}
\newcommand{\SSg}{\tilde{\Sigma}}
\newcommand{\eq}{\begin{equation}}
\newcommand{\eqx}{\end{equation}}
\newcommand{\eqn}{\begin{eqnarray}}
\newcommand{\eqnx}{\end{eqnarray}}
\newcommand{\ben}{\begin{eqnaray}}
\newcommand{\een}{\end{eqnarray}}
\newcommand{\f}[2]{\frac{#1}{#2}}
\newcommand{\ra}{\longrightarrow}
\newcommand{\GG}{{\cal G}}
\renewcommand{\AA}{{\cal A}}
\newcommand{\GR}{G(z)}
 \newcommand{\MM}{{\cal M}}
\newcommand{\BB}{{\cal B}}
\newcommand{\ZZ}{{\cal Z}}
\newcommand{\DD}{{\cal D}}
\newcommand{\HH}{{\cal H}}
\newcommand{\RR}{{\cal R}}
\newcommand{\VV}{V_A}                
\newcommand{\GT}{{\cal G}_1 \otimes {\cal G}_2^T}
\newcommand{\GGb}{\bar{{\cal G}}^T}
\newcommand{\Du}{{\cal D}_1}
\newcommand{\Dl}{{\cal D}_2}
\newcommand{\zb}{\bar{z}}
\newcommand{\trqq}{\tr_{q\bar{q}}}
\newcommand{\arr}[4]{%
\left(\begin{array}{cc}%
#1&#2\\
#3&#4
\end{array}\right)
}
\newcommand{\arrd}[3]{
\left(\begin{array}{ccc}
#1&0&0\\
0&#2&0\\
0&0&#3
\end{array}\right)
}
\newcommand{\tr}{\mbox{\rm tr}\,}
\newcommand{\trn}{\mbox{\rm tr}_N\,}
\newcommand{\One}{\mbox{\bf 1}}
\newcommand{\pauli}{\sg_2}
\newcommand{\corr}[1]{\la{#1}\rangle}
\newcommand{\br}[1]{\overline{#1}}
\newcommand{\phib}{\br{\phi}}
\newcommand{\psib}{\br{\psi}}
\newcommand{\lm}{\lambda}
\newcommand{\ksi}{\xi}
\newcommand{\Gb}{\br{G}}
\newcommand{\Vb}{\br{V}}
\newcommand{\Gm}{G_{q\br{q}}}
\newcommand{\Vm}{V_{q\br{q}}}
\newcommand{\ggd}[2]{\GG_{#1}\otimes\GG^T_{#2}\Gamma}
\newcommand{\noi}{\noindent}

\newcommand{\refnote}[1]{#1}
	
\vspace*{12mm}
\begin{center}
 \begingroup
   \def\thefootnote{\fnsymbol{footnote}}%
  \thispagestyle{empty}\let\thanks\relax
{\titlefont VARIOUS SHADES OF BLUE'S FUNCTIONS\footnote{Lectures  presented by
MAN at the XXXVII Cracow School of Theoretical Physics, Zakopane, 1997.}}
 \endgroup
 \setcounter{footnote}{0}%
\end{center}
 \vspace*{12mm}
\begin{center}
{\appauthcs Romuald A. Janik$^1$, Maciej A. Nowak$^{1,2,3}$,
G\'{a}bor Papp$^{2,4}$, and \\ Ismail Zahed$^{5}$}
\end{center}

\begin{center}
{\small\rm $^1$Department of Physics, Jagellonian University,
30-059 Krak\'{o}w, Poland\\
$^2$GSI, Planckstr. 1, D-64291 Darmstadt, Germany\\
$^3$Institut f\"{u}r Kernphysik, TU Darmstadt, D-64289
Darmstadt,
Germany\\
$^4$Institute for Theoretical Physics, E\"{o}tv\"{o}s
University,
Budapest, Hungary\\
$^5$Department of Physics, SUNY, Stony Brook, NY 11794, USA}
\end{center}

\vspace*{20mm}
{\small\rm
We discuss random matrix models in terms of elementary operations on
Blue's functions (functional inverse of Green's functions).
We show that such operations embody the essence of a number of physical
phenomena whether at/or away from the critical points. We
illustrate these assertions by borrowing on a number of recent results
in effective QCD in vacuum and matter. We provide simple physical
arguments in favor of the universality of the continuum QCD spectral
oscillations, whether at zero virtuality, in the bulk of the spectrum or
at the chiral critical points. We also discuss effective quantum systems
of disorder with strong or weak dissipation (Hatano-Nelson localization).
}

\vfill\eject
\tableofcontents
\vfill\eject

\section{INTRODUCTION}

Complex and disordered systems play an important role
in a number of areas in physics ranging from mesoscopic conductance
fluctuations to simplicial quantum gravity. The dual interplay between 
disorder and localization~\cite{ANDERSON}, and disorder and chaos
\cite{BOHIGAS}, has far reaching consequences on many aspects
of modern physics whether at the macroscopic or microscopic level. 

An important tool for investigating Hamiltonians of disordered systems is
random matrix theory~\cite{WIGNER,PORTER,DYSON,MEHTA}. A number of
techniques have been developed such as the method of orthogonal
polynomials, the supersymmetric and replica methods to cite afew.
A powerful alternative to some of these methods has been the work of
Voiculescu~\cite{VOICULESCU} in the context of operator algebras and
the ensuing concept of free random variables. Free disorder allows for 
an exact calculation of various moments of a Hamiltonian composed
of random plus deterministic parts, hence it's Green's function,
while maintaining its parts logically separated. In physical systems, 
freeness amounts to summing over all `single-site' rescatterings~\cite{HNUS},
a procedure that is close to mean-field treatments in many-body physics.
In systems amenable to a $1/N$ expansion, this procedure is equivalent 
to the one based on the concept of independent disorder in the large $N$
limit. In most cases, it is simpler to implement and holds for arbitrary
forms of disorder.

In these lectures we will review and generalize the basic operations 
introduced by Voiculescu for free random variables, and show how they
could be used in a variety of physical applications. Instead of using
the standard language of Green's functions, we choose to work with 
their functional inverses introduced and named  Blue's functions 
by Zee \cite{ZEE}.
We show that the use of Blue's functions sheds new light on a number 
of issues in conventional as well as non-conventional random matrix 
theory, while providing an effective calculational shortcut. In many ways,
it is the user-friendly alternative to Voiculescu's involved concepts of 
non-commutative independence~\cite{VOICULESCU}  and non-crossing
cumulants to cite a few~\cite{SPEICHERCUM}. The 
concept of addition law between Blue's functions in random matrix theory 
is perhaps the closest in spirit to the concept of a mean-field analysis 
in many-body physics. Indeed, for quantum problems with diagonal disorder
it is equivalent to the Coherent Phase Approximation (see below 
and~\cite{NEUSPEICHER}).

In the first part of these lectures we will introduce the elementary 
calculus with Blue's functions. We will define the concepts of {\it 
addition law} and  {\it multiplication law } for Blue's functions 
for hermitian random ensembles with arbitrary
measures, and the concept of generalized {\it addition law} 
for generalized Blue's functions for nonhermitian random ensembles
with arbitrary measure. 
 We also define the concepts of {\it differentiation} and
{\it integration} of Blue's functions and their respective relationship
with the end-points and boundaries of eigenvalue spectra. 
Finally we exploit some {\it analytical properties} of Blue's functions
by constructing conformal mappings between the domains of analyticity
for hermitian and non-hermitian ensembles. 
In the second
part of these lectures, we will present a number of applications to 
effective Quantum Chromodynamics (QCD), and models of quantum mechanics 
with strong and weak dissipation. 

In effective QCD we emphasize the role played by the quark modes near
zero virtuality, in the characterization of the issue of spontaneous
symmetry breaking in the vacuum and near the chiral critical points. We
present simple arguments for the importance of chiral symmetry in the
conditioning of the spectral fluctuations near zero virtuality as well
as within the bulk of the low-lying eigenvalue spectrum, both in vacuum
and matter. We introduce simple examples of chiral random matrix models,
and provide plausible arguments for the decoupling of the soft and hard
modes in infrared sensitive observables. We give a schematic description
of the U(1) problem and its resolution, and illustrate the ingredients
at work in the screening of the topological charge. We also show how the
concepts of temperature and chemical potential may be {schematically}
incorporated in the present approach, and how the structural changes in
the eigenvalue spectra are encoded in the Blue's functions. We briefly
comment on the role of masses and vacuum angle $\theta$ in this
effective approach to QCD. In the models of quantum mechanics, we
demonstrate the versatility of our approach to the problem of open
chaotic scattering as well as tunneling in the presence of
dissipation. We go on showing that the addition law for Blue's functions
for a class of quantum systems with dissipation is the analogue of the
Coherent Phase Approximation (CPA), a method commonly used for the
Anderson model \cite{ANDERSON} and its
variants~\cite{NEUSPEICHER,WEGNER}.  We speculate on further
applications of this approach in our conclusions.


\newcommand{\sg}{\sigma}
\newcommand{\al}{\alpha}
\newcommand{\PP}{{\cal P}}
\newcommand{\dz}{\partial_z}
\newcommand{\dzb}{\partial_{\zb}}
\newcommand{\ggb}{\GG\otimes \br{\GG}^T \cdot \Gamma}
\newcommand{\gbgb}{\br{\GG}\otimes \br{\GG}^T \cdot \Gamma}
\newcommand{\ggg}{\det(1-\GG\otimes \br{\GG}^T \cdot \Gamma)}

\section{OPERATIONS on BLUE'S FUNCTIONS}

The fundamental problem in random matrix theories
\cite{PORTER,MEHTA,WEIDNEW,ZINNJUSTIN} is to find the 
distribution of eigenvalues in the large $N$ (size of the matrix $\MM$)
limit. One encounters two generic situations --- in the case of
hermitian matrices the eigenvalues lie on one or more intervals on the real
axis, while for general non-hermitian ones, the eigenvalues occupy
two-dimensional domains (or even fractals) in the complex plane. In both
cases the distribution of eigenvalues 
can be reconstructed from the knowledge of the Green's function
\be
G(z)=\frac{1}{N}\left< {\rm  tr} \frac{1}{z-\MM}\right>
\label{green}
\ee
where averaging is done over the ensemble of $N \times N$ random matrices
generated with probability
\be
P(\MM)=\frac{1}{Z}e^{-N {\rm tr } {\cal V}(\MM)}\,.
\ee
For hermitian matrices the discontinuities of the Green's function
coincide with the support of eigenvalues.
In the simplest example of the Gaussian potential 
${\cal V}(\MM)= \f{1}{2}\MM^2$ for a 
{\it hermitian} matrix $\MM$, a standard calculation gives~\cite{BZ}
\be
G(z)=\frac{1}{2}(z-\sqrt{z^2-4})\,.
\label{gauss}
\ee
The reconstruction of the spectral function (eigenvalue distribution)
is based on the well known relation
\be
\frac{1}{x\pm i\epsilon}= {\rm PV} \frac{1}{x} \mp i\pi \delta(x)
\ee
hence
\be
\nu(\lambda)= -\frac{1}{\pi} \lim_{\epsilon \rightarrow 0} {\rm Im} 
G(z)|_{z=\lambda +i\epsilon} \,.
\label{CUTX}
\ee
In the Gaussian case (\ref{gauss}), the discontinuities in the 
Green's function come from the cut of the square root, leading to 
Wigner's semicircle law $2\pi\nu(\lambda)=\sqrt{4-\lambda^2}$
for the eigenvalue distribution.
In practice, the averaging procedure $<...>$ for more complicated 
probability distributions is difficult. It could be simplified if 
the randomness is assumed to  be free (see below).

In the general case of non-hermitian matrices, the two-dimensional
domains where the eigenvalues are distributed are exactly those where
the Green's functions {\it cease} to be holomorphic, that is
\be 
\nu(z,\bar{z})=\frac{1}{\pi} \partial_{\bar{z}}  G (z,
\bar{z}) \neq 0 \,.
\ee

\subsection{Addition}

Let us consider the problem of calculating Green's function 
for the sum of two independent ensembles $\MM_1$ and $\MM_2$, 
i.e.
\eqn
G(z)&=&
\frac{1}{N} \int [d\MM_1][d\MM_2] P_1(\MM_1)P_2(\MM_2)
\ {\rm tr} \frac{1}{z-\MM_1-\MM_2}\nonumber\\ 
&\equiv&\frac{1}{N}\left< {\rm tr} \frac{1}{z-\MM_1-\MM_2}\right> \,.
\label{conv}
\eqnx
The concept of addition law for hermitian ensembles 
was introduced by Voiculescu \refnote{\cite{VOICULESCU}}. 
In brief, he
proposed an additive transformation (R transformation), which linearizes
the convolution of non-commutative matrices (\ref{conv}), much like the 
logarithm of the Fourier transform for the convolution 
of  arbitrary functions. This method is 
 an important shortcut to obtain the equations for the Green's
functions
for a sum of matrices, starting from the knowledge of the Green's functions
of individual ensembles of matrices. This formalism was reinterpreted
diagrammatically by 
Zee~\refnote{\cite{ZEE}}, who introduced the concept of Blue's function.

Let us consider the problem of finding the Green's function 
of a sum of two  independent (free~\refnote{\cite{VOICULESCU}})
 random matrices ${\cal M}_1$ and ${\cal
M}_2$, provided we know the Green's functions of each of them 
separately. The key observation is that the 
 self-energy $\Sigma$, defined as $\Sigma(z)=z-G^{-1}(z)$,
 can be always expressed
as a function of $G$ itself and {\em not  z} as usually done in 
textbooks. Then the following addition law holds
\be
 \Sigma_{1+2}(G)= \Sigma_1(G) + \Sigma_1(G) .
\label{sumgg}
\ee
This formula does not hold for $G\rightarrow z$.
Voiculescu's R transform is nothing but the
self-energy expressed as a function of $G$ 
({\it i.e.} $R(G)\equiv \Sigma(G)$).
For the Gaussian randomness $R(u)=u$. For an arbitrary complex number $u$,
the addition (\ref{sumgg}) reads $R_{1+2}(u)=R_1(u)+R_2(u)$. The set of all
R transforms generate an Abelian group. The Blue's function, introduced by 
Zee~\refnote{\cite{ZEE}}, is simply 
\be
B(G)=\Sigma(G)+G^{-1} .
\label{blue}
\ee
Therefore, using the identity $G(z)=(z-\Sigma)^{-1}$, we see that the
Blue's function is the functional inverse of the Green's function
\be
B[G(z)]=z
\label{blueinv}
\ee
and the addition law for Blue's functions reads
\be
B_{1+2}(z)=B_1(z)+B_2(z)-\frac{1}{z} .
\label{addblue}
\ee
Using the definition of the Blue's function we could rewrite the last
equation in ``operational'' form as
\be
z=B_1(G)+B_2(G)-G^{-1}
\label{addbluebis}
\ee
with $G_{1+2}\equiv G$. 
The algorithm of addition is now surprisingly simple~\refnote{\cite{ZEE}}:
Knowing $G_1$ and $G_2$, we find $B_1$ and $B_2$ from (\ref{blueinv}),
and read from (\ref{addbluebis}) the final equation for the resolvent 
of the sum. Note that the method treats on equal footing the Gaussian and
non-Gaussian ensembles, provided that the measures $P_1$ and $P_2$ are 
independent (free).

The generalization of this algorithm to the case of arbitrary complex
random matrices
is subtle.  In this case, the eigenvalues are complex  and are 
not distributed along the 
one dimensional cuts on the real axes, but rather form on the $z$ plane 
two-dimensional surfaces or fractals \cite{DERRIDA}. Again we would like 
to have a method for finding the Green's function (\ref{green}).
 
The generalization~\refnote{\cite{DIAG}} amounts to consider,
 in addition to the original
matrix $\MM$, a hermitian conjugate copy $\MM^\dagger$ and an
infinitesimal `coupling' $\lambda$ between both copies.
This leads us to investigate the matrix-valued Green's function
\be
\hat{{\cal {G}}}=\arr{{\cal G}_{qq}}{{\cal G}_{q\overline{q}}}%
 {{\cal G}_{\overline{q}q}}
 {{\cal G}_{\overline{q}\overline{q}}}_{2N\times 2N}
= \left\langle \arr{z-\MM}{\lambda}{\lambda}{\zb 
 -\MM^{\dagger}}^{-1}_{2N\times 2N}\right\rangle \,.
\label{19}
\ee
The subscript $2N\times 2N$ indicates that we are dealing with a $2N$
by $2N$ matrix --- $\MM$ and $\MM^\dagger$ are $N$ by $N$ matrices,
and $z$ and $\lambda$ are implicitly multiplied by an $N\times N$
identity matrix. 
A variant of this   construction leading to similar results was
recently proposed  in \cite{ZEENEW1}. 
Now, the ordinary Green's function (\ref{green}),
the chief aim of the calculation, is just the trace
of the upper left-hand corner of $\hat{{\cal{G}}}$:
\be
G(z)=\lim_{\stackrel{N\rightarrow\infty}{\lambda\rightarrow0}}\f{1}{N} 
 \trn {\cal G}_{qq}
\ee 
where the limit $N\rightarrow\infty$ is understood before $\lambda\rightarrow 
0$. However in ordinary calculations one never has to perform any
explicit limits. 
It is convenient to take the traces of the four $N$ by $N$ blocks of
$\hat{{\cal G}}$ to obtain a $2\times 2$ matrix:
\be
\GG\equiv \arr{G_{qq}}{G_{q\overline{q}}}{G_{\overline{q}q}}
{G_{\overline{q}\overline{q}}}_{2\times 2} \equiv
 \arr{\trn{\cal G}_{qq}}{\trn{\cal G}_{q\overline{q}}}%
  {\trn{\cal G}_{\overline{q}q}}
{\trn{\cal G}_{\overline{q}\overline{q}}}_{2\times 2} \,.
\ee

The generalized Blue's function~\refnote{\cite{DIAG,NONHER}}
 is now a matrix valued function of
a $2 \times 2$ matrix variable defined by
\be
\BB(\GG)=\ZZ=\left( \begin{array}{cc} z & \lambda \\ \lambda & \bar{z}
                    \end{array} \right) 
\ee
where $\lambda$ will be eventually set to zero. This is equivalent
to the definition in terms of the self-energy matrix
\be
\BB (\GG )=\Sigma+\GG^{-1}
\label{genblue}
\ee
where $\Sigma$ is a $2 \times 2$ self energy matrix expressed as a
function of a matrix valued Green's function.    
The generalized addition law reads
\be
\ZZ  =\BB_1(\GG)+\BB_2(\GG) -\GG^{-1} .
\label{genadd}
\ee
This matrix equation has always two kinds of solutions. 
The first one (trivial) corresponds to the case when the
off-diagonal matrix elements
of $\GG$ are equal to zero. In this case, (\ref{genadd}) splits 
into two copies of the addition law (\ref{addblue}), for the 
holomorphic function 
$B(z)$ and its anti-holomorphic copy $B(\bar{z})$, respectively. 
The second (non-trivial)  solution corresponds to the case
when holomorphic separability of (\ref{genadd}) is no longer possible, 
and all matrix elements of $\GG$ are nonholomorphic functions of 
$z$ {\it and} $\bar{z}$.  The average 
density of eigenvalues follows then from the
electrostatic analogy to two-dimensional Gauss law~\refnote{\cite{SOMMERS88}} 
\be 
\nu(z,\bar{z})=\frac{1}{\pi} \partial_{\bar{z}}  G_{qq} (z,
\bar{z}) \,.
\ee
The condition that the holomorphic and nonholomorphic solutions match
(i.e. $G_{q\bar{q}}(z,\bar{z})=0$) defines the borderline
of the eigenvalue distribution of the $z$ plane.
The advantage of the addition law,  for hermitian and non-hermitian 
random matrix models, stems from the fact that it
treats Gaussian and non-Gaussian randomness on the same 
footing~\refnote{\cite{ZEENEW2}}.

\subsection{Multiplication}
We now demonstrate that Blue's functions provide also an important shortcut
to obtain the equation for the Green's (Blue's) function 
for a product of matrices, starting from the knowledge of Green's (Blue's) 
functions of individual ensembles of matrices, i.e. to find  
\be
G(z)=\frac{1}{N}\left< {\rm tr} \frac{1}{z-\MM_1 \cdot \MM_2}\right>   
\label{product}
\ee
provided that $\MM_1$ and $\MM_2$ are free and 
$\langle{\rm tr}\MM_1\rangle, \langle{\rm tr}\MM_2\rangle\neq 0$. 
Let us introduce the notation for continued fraction
\be
\frac{w}{\Sigma_{\MM}\left( \frac{w}{\Sigma_{\MM}\left(
 \frac{w}{\Sigma_{\MM}(...)}\right)}\right)}  
\equiv   \frac{w}{\Sigma_{\MM}(\bullet)} \,.
\label{bullet}
\ee
Then the multiplication law for Blue's functions reads
\be
B_{1\star
2}\left(\frac{w}{\Sigma_{1\star2}(\bullet)}\right)
=
\frac{w}{1+w}
B_1\left(\frac{w}{\Sigma_1(\bullet)}\right) \cdot 
B_2\left(\frac{w}{\Sigma_2(\bullet)}\right) \,.
\label{prodblue}
\ee
For the diagrammatic proof of this relation we refer to 
\refnote{\cite{JANIKPHD}}. Contrary to standard Green's functions,
the Blue's functions provides the factorization mechanism for the averaging 
procedure in (\ref{product}). 
Using the definition $B(G)=\Sigma(G)+G^{-1}$
we get $B(w/\Sigma(\bullet))=\Sigma(\bullet)(1+1/w)$ therefore
the multiplication law reads
\be
\Sigma_{1\star
2}\left(\frac{w}{\Sigma_{1\star2}(\bullet)}\right)
=
\Sigma_1\left(\frac{w}{\Sigma_1(\bullet)}\right) 
\cdot \Sigma_2\left(\frac{w}{\Sigma_2(\bullet)}\right) \,.
\label{prodsigma}
\ee
Note that the S-transformation of Voiculescu is related to 
$\Sigma(\bullet)$ via~\refnote{\cite{JANIKPHD}}
 \be 
S(w)=\frac{1}{\Sigma\left(\frac{w}{\Sigma(\bullet)}\right)}
\label{relation}
\ee
hence (\ref{prodsigma}) reads $S_{1\star2}(w)=S_1(w)\cdot S_2(w)$.

To check that the relation (\ref{relation}) with Voiculescu's
S-transform is true, we have to check that the function $\chi(w)$
defined by $S(w)=\chi(w)(1+1/w)$ satisfies~\cite{VOICULESCU}
\eq
\chi\big(zG(z)-1\big)=1/z \,.
\eqx
Comparing the definition of $\chi$ and (\ref{relation}) we see that
indeed
\be
\chi(w)=\frac{1}{(1+1/w)\Sigma\left(\frac{w}{\Sigma(\bullet)}\right) }
=\frac{1}{B(w/\Sigma)} = \frac{1}{B[G(z)]} = \frac{1}{z} 
\ee
where in the last two equalities we changed the variables $w=zG(z)-1$ and 
used the definition of Blue's function, respectively.     

\subsection{Differentiation}

In the previous two sections we showed how to use Blue's functions to
obtain the Green's function of a sum or product of random matrices
from the properties of the individual ensembles. Now we will
concentrate on the information about the structure of the eigenvalues
of some fixed random matrix ensemble, which we may extract from the
Blue's function.

As shown by Zee~\cite{ZEE}, the differentiation of the Blue's function 
leads to the determination of the endpoints of the eigenvalue distribution 
for hermitian random matrices. 
One could easily visualize this point recalling the form of the Green's
function for the Gaussian distribution (\ref{gauss}). When approaching the
endpoints
$z=\pm 2$, the imaginary part of the Green's function vanishes like
$\sqrt{(2-\lambda)(2+\lambda)}$. Therefore, the endpoints fulfill the equation
$G'(z)|_{z=\pm 2} =\infty$, which may be used as the 
defining equation for {\it unknown} 
endpoints in the case of more complicated ensembles, since the
endpoints are the branching points of the resolvent.
Since Blue's function is the functional inverse of the Green's function,
the  location of the endpoints could be inferred from its extrema
\be
\frac{d B(G)}{dG} =0 \,.
\label{blueend}
\ee

\subsection{Integration}

In this section we will consider the phase structure of a class of
systems with a partition function depending on a parameter $z$:
\eq
Z_N (z) =\corr{\det(z-\MM)} \,.
\label{zpart}
\eqx
In general $\MM$ may be a sum of a random matrix and some
deterministic one (not necessarily hermitian). The variable $z$ may also be
promoted to a matrix. A typical example being the chiral random matrix model 
for effective QCD as discussed below. Generically, the analytical
structure of 
(\ref{zpart}) and its zeros are of interest for phase studies in generalized 
statistical mechanics \cite{HUANG}. In matrix models, this information is all
contained in the Blue's functions.

For finite $N$ the partition function  is a polynomial in $z$ and has
$N$ (Yang-Lee) zeroes
\eq
Z_N (z) =(z-z_1)(z-z_2)\dots (z-z_N) \,.
\eqx
Taking the logarithm and approximating the density of zeroes by a
continuous distribution we get
\eq
\log\cor{\det(z-\MM)} = \sum_i \log(z-z_i) = \int \rho(z') \log(z-z') 
	dz' \,.
\label{INTEGRAL}
\eqx
We see that the density of  zeroes can be reconstructed from
the discontinuities of the {\em unquenched} Green's function
\eq
\label{der}
\partial_z \log\cor{\det(z-\MM)}=\cor{\tr \f{1}{z-\MM} \cdot \det (z-\MM)}
\eqx
through Gauss law~\refnote~{\cite{VINK}}. For $z$ close to infinity the
unquenched and quenched (\ref{green}) Green's functions
coincide\footnote{In leading order in $1/N$.}. Moreover the unquenched
resolvent is nonsingular configuration by configuration, hence
holomorphic.
Therefore it is exactly the functional inverse of the ordinary
holomorphic Blue's function. In general we may have several branches for
the 
Green's functions, which are determined by different saddle points
contributing to (\ref{zpart}). The discontinuity (cusp) and hence the
location of the zeroes is determined by the possible contribution
of two of them. From a $1/N$ expansion
\eq
\log Z_N=N E_0+E_1+\f{1}{N}E_2+\ldots
\label{YL1}
\eqx
we see that two saddle points may contribute if
\eq
{\rm Re} E_0^{sp.I}={\rm Re} E_0^{sp.II}
\label{cuspline}
\eqx
and $E_0$ is determined by (\ref{der})
\eq
E_0=\int^z dz' G(z')+const
\label{blueint}
\eqx
or equivalently
\be
E_0 =zG - \int\! dG\ B(G) + const
\label{trick}
\ee
after integrating by part. Note that we have used the fact 
that $z(G) =B(G)$ is just the Blue's 
function of $G$.
The constant in $E_0$ is fixed by the 
asymptotic behavior of (\ref{zpart}), that is $Z_N\sim z^N$.
We should add that performing the integration using Blue's function is
in most cases much simpler than the corresponding direct integration
of $G(z)$.

If we denote  the  complex valued set of  all the cusps 
by 
\eq
F(z,\bar{z}) \equiv {\rm Re} E_0^{sp.I}-{\rm Re} E_0^{sp.II} =0
\eqx 
then, to leading order, the distribution of 
singularities along the ``cusps" (\ref{cuspline}) is 
\be
\varrho (z) = \frac 1{2\pi} | \dz F |^2 \,\, \delta ( F (z, \zb) )
\label{densdel}
\ee
which is normalized to 1 in the $z$-plane. Redefining the density of 
singularities by unit length along the curve $F(z,\zb)=0$, we may rewrite
(~\ref{densdel}) as
\be
\left. \varrho(z)\right|_{F=0} = \frac 1{2\pi} |\dz F | \equiv
	\frac 1{4\pi} |G^{(i)}-G^{(j)}| \,.
\label{densun}
\ee
For $\varrho \neq 0$, the integrand
in (\ref{INTEGRAL}) is singular at $z=z'$ which results in
 a cusp. For $\varrho =0$, that is $\dz F =0$,
(\ref{INTEGRAL}) is differentiable. For physical 
$E_0$ (real and monotonically increasing), the points $\varrho =0$ are 
multi-critical points. At these points all $n$-point ($n\geq 2$)
functions diverge.
This observation also holds for Ising models with complex external
parameters \refnote{\cite{ROBERT}}. Assuming
macroscopic
universality~\refnote{\cite{AMBJORN,MAKEENKO,AKEMANN,AMBJORNETAL,BREZIN}}
for all $n$-point functions($n\geq 2$),  
we conclude that $\partial_z F =0$ means a branching point for the
resolvents, hence $\dz G =\infty$ 
or $B' (G) =0$, as shown before.

The next to leading contribution $E_1$ in (\ref{YL1}) does not influence the
preceding discussion except at exceptional points\footnote{For a detailed 
discussion of this point see \protect{\refnote{\cite{DIAG}}}.}. In fact 
the expression for $Z_N$ is
\be
Z_N (z) =e^{N E_0}\cdot \left(\left\{ \det C(G)\right\}^{-\frac 12} + 
{\cal O} \left( \frac 1N \right) \right) \,.
\label{X107}
\ee
The explicit form of the 
argument of the determinant could be addressed using several methods.
We do not discuss this point here. We would like to stress, that 
usually $C$ is a simple function of the resolvent $G$ on a specific branch.
In such a way the Blue's function conditions the leading behavior 
of the partition functions, either explicitly via integration leading to 
$E_0$, or implicitly via the functional inverse and the macroscopic 
universality of higher correlators. 

Note that due to the power $-1/2$ in (\ref{X107}), 
the fluctuations have ``bosonic" character, and are dwarfed  
by the ``quark-loop" contribution in the ratio $1:N$ \refnote{\cite{USNJL}}.
In particular, the condition for the divergence of the ``bosonic'' 
fluctuation 
at $C(G)=0$, signals the breakdown of the $1/N$ expansion. These are the 
exceptional points, for which new regimes of microscopic universality may
set in, following pertinent scalings. The standard example for Gaussian 
randomness is the divergence of the wide two-point correlator
for $z^2-4=0$, signaling the universal behavior (Airy oscillations)
\cite{AIRY} near the endpoints of the spectra. In the case of chiral 
Gaussian randomness the divergence happens for $z(z^2-4)=0$. The additional 
multicritical point $z=0$ is due to the chiral nature (``Goldstone pole'') 
and signals another class of universal behavior (Bessel oscillations) 
\cite{BESSEL}.

\renewcommand{\dz}{\partial_z}
\renewcommand{\dzb}{\partial_{\zb}}
\newcommand{\ddx}{\partial_x}
\newcommand{\ddy}{\partial_y}

\subsection{Mapping}

The concept of Blue's functions has a very surprising interpretation
in terms of conformal mappings between two distinct ensembles of random
matrices. Indeed, these mappings can take the discontinuities or cuts of
the eigenvalue distribution of a hermitian random matrix model and transform 
them into the boundary of the eigenvalue distribution of its non-hermitian 
analogue \cite{NONHER}. These transformations can be carried fully in the 
holomorphic region, thereby avoiding the subtle issue of nonholomorphy
altogether.


To illustrate these points, consider the case  where a Gaussian random 
and hermitian matrix $H$ is added to an arbitrary deterministic matrix $M$. 
The addition law for Blue's functions states that
\be
B_{H+M}(u)=B_H(u)+B_M(u)-\frac{1}{u}= u+B_M(u)
\label{hm}
\ee
where we have used  explicitly that for a Gaussian ensemble
$B_H(u)=u + 1/u$.
 Now, if we were to note that in the {\em holomorphic} domain
the Blue's  transformation for the Gaussian  anti-hermitian ensemble
is $B_{iH}(u)=-u + 1/u$,
we find\footnote{Note that in the holomorphic domain, 
anti-hermitian Gaussian nullifies the hermitian Gaussian 
in the sense of group property for addition law for the $R$ transformation.}
\be
B_{iH+M}(u)=B_{iH}(u)+B_M(u)-\frac{1}{u}=-u+B_M(u).
\label{ihm}
\ee
These two equations yield
\be
B_{iH+M}(u)=B_{H+M}(u)-2u \,.
\label{sumBLet}
\ee
Substituting $u\rightarrow G_{H+M}(z)$ we can rewrite (\ref{sumBLet}) as
\be
B_{iH+M}[G_{H+M}(z)]=z-2G_{H+M}(z) \,.
\label{map1}
\ee
Let $w$ be a point in the complex plane for which
$G_{iH+M}(w)=G_{H+M}(z)$.
Then, using the definition of the Blue's function, we get 
\be 
w=z-2G_{H+M}(z)\,.
\label{map2}
\ee
The result (\ref{map2}) provides a conformal transformation mapping
the {\it holomorphic} domain of the ensemble $H+M$ ({\it i.e.}
the complex plane $z$ minus cuts) onto the {\it holomorphic} domains of 
the ensemble $iH+M$, ({\it i.e.} the complex plane $w$ minus the 
``islands''), defining in this way the support of the eigenvalues.

\section{UNIVERSALITY without BLUE'S FUNCTIONS}

The concept of Blue's functions rely on the use of the $1/N$ expansion.
There is a number of circumstances where the standard $1/N$ expansion
breaks down, and where interesting phenomena take place in spectral
analysis. In this part of our lectures, we will discuss them in the
context of effective QCD.  The Euclidean Dirac spectrum in effective QCD
presents the global structure described qualitatively in
Fig.~\ref{qcdspectrum}. For gauge configurations with non-zero winding
number $n\neq 0$, an accumulation of zero modes take place at zero as
expected from the Atiyah-Singer index theorem. The large eigenvalues
$\lambda$ do not sense sufficiently localized gauge configurations. They
are described by a free density of states $\nu(\lambda )\sim
V|\lambda|^3$ with $V$ the Euclidean four-volume. Near zero, the density
of states is conditioned by the spontaneous breaking of chiral
symmetry. This phenomenon causes a huge accumulation of eigenvalues,
with a level spacing typically of order $1/V$. In this regime the $1/N$
expansion has to be reorganized to accomodate for new scaling laws. We
now proceed to describe some of these features and other related
properties, using effective QCD.
\begin{figure}
\centerline{%
  {\bf a.)}{\epsfysize=30mm \epsfbox{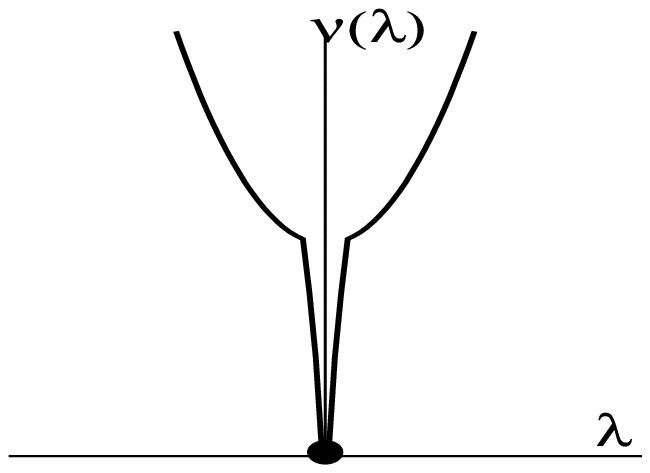}} \hfill
  {\bf b.)} {\epsfysize=30mm \epsfbox{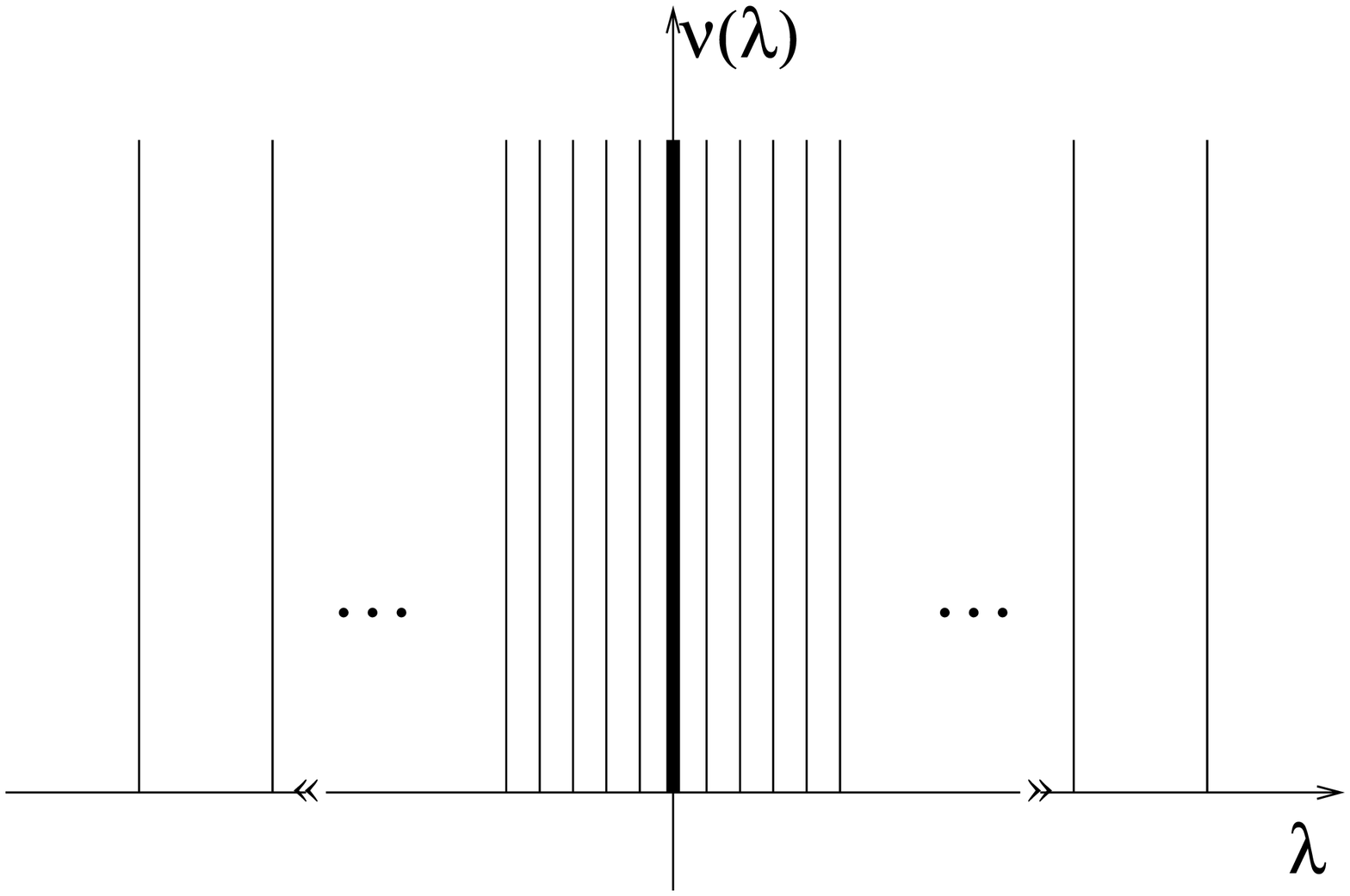}}%
}
\caption{Typical Dirac Spectrum for massless chiral quarks in Euclidean space
(a). The enlargement around zero virtuality for finite volume is shown
in (b).}
\label{qcdspectrum}
\end{figure}

\subsection{Effective QCD and Broken Chiral Symmetry}

Of all the effective model approaches to QCD, most noteworthy are those
that account properly for the {\it spontaneous} breaking of chiral
symmetry~\cite{BOOK}. There are a number of effective models in this
direction, such as the linear and non-linear sigma models and their
progenies, the Nambu-Jona-Lasinio model and its variants~\cite{NAMBU},
the instanton model~\cite{BOOK,INSREV}, and more recently the chiral
random matrix models.\footnote{We note at this stage that chiral
perturbation theory \cite{GASSER} or chiral reduction
formulae~\cite{YAMZAH}, are not models to QCD but a way to analyze its
chiral content using (broken) chiral symmetry and data.} Since these
lectures are devoted to random matrix models, we will focus our
attention on the latters.

Why do chiral random matrix models have anything to do with effective
QCD? The answer to this question is simple: in the long wavelength limit, 
chiral symmetry is spontaneously broken in the vacuum. As a result the
symmetry of the associated coset space as well as the mode of 
explicit chiral breaking determines uniquely the character of the
QCD effective action in the infinite wavelength limit and to leading
order in the symmetry breaking.

For $N_f>1$ the coset space
is isomorphic to $SU(N_f)$ and the mode of explicit breaking is 
in the fundamental $(N_f,N_f)$ representation (ignoring the 
induced U(1) breaking from the anomaly~\footnote{The inclusion of the
U(1) anomaly upsets the chiral power counting, except in large $N_c$
where the anomaly is treated perturbatively~\cite{WITTEN,SMILGA}}). The
partition function of four dimensional QCD on an Euclidean hypertorus of
volume $V$ with finite vacuum angle $\theta$ is \cite{GASSER,SMILGA}
\be
Z (\theta, M ) = Z_0 \int\,\, [d U] 
e^{-\frac 12 V\Sigma\,\, {\rm Tr} \left( (M e^{i\theta/N_f} U + 
U^{\dagger} M \, e^{-i\theta /N_f} )
+{\cal O} \right)}  
\label{EFF1}
\ee
where the mass matrix $M$ is real, positive and diagonal, and $U$ is 
$SU(N_f)$ valued. ${\cal O}$ are the nonzero mode contributions 
as well as terms of order $M^2$ and higher. $Z_0$ is an overall $M,\theta$ 
independent normalization,
and $V$ is the space-time volume. $\Sigma= |\left< \bar{q} q \right> |$ 
is a measure of the vacuum  condensate in the chiral limit with a 
normalization scale matching that of the quark mass matrix
\footnote{For $N_f=1$ chiral symmetry is explicitly broken by the U(1)
anomaly. Nevertheless, power counting in $M$ is still valid and 
the representation (\ref{EFF1}) on the U(1) coset still holds, provided that
the term $V\chi_* (i{\rm ln\,\,det\,\,}U)^2$ is added \cite{WITTEN}.}. Chiral
random matrix models share (\ref{EFF1}) with 
QCD~\cite{USPAST,SHURYAKVER,BESSEL}.
(This is also the case of all the other models mentioned above.)

In the limit $VM\Sigma\gg 1$ (macroscopic regime), the integrand in 
(\ref{EFF1}) is peaked. As a result, the contribution is dominated by the 
saddle point with a preferred direction on the coset, say $U\sim {\bf 1}$.
In this regime, chiral symmetry is spontaneously broken, with a vacuum
condensate of the order of $\Sigma$ (in absolute value). The saddle point
contribution is extensive with a vacuum energy density ${\cal E} = M\Sigma 
+{\cal O}$. The corrections ${\cal O}$ may be organized in chiral power 
counting $M\sqrt{V}\sim 1$. The additional tree contribution to the
saddle point due to the omitted non-zero modes is of order $1/\sqrt{V}$,
and vanishes in the large volume limit. Similarly, the loop contribution
from the non-zero modes (massive bosons) as well as the counterterms to
order $1/V$ vanishes as $V\rightarrow\infty$. The former disappears
exponentially through Boltzman-like factors
$e^{-\beta\sqrt{M}}$ with $\beta^4=V$. The divergences in the loops are solely
extensive (same in small and large volumes) and readily absorbed in the 
renormalization of the cosmological constant. Chiral random matrix models
support this rationale only to tree level in the macroscopic limit, and
hence should be regarded as effective models in this regime.

In the limit $VM\Sigma\ll 1$  (microscopic regime), 
the condensate is of the order of the volume $V$,
and vanishes in the chiral limit (Mermin-Wagner-Coleman theorem). 
The calculation can still be organized in powers of $M$ and $1/V$ 
such as $MV\sim 1$\footnote{This is just a reorganization of the chiral
power counting.}. In this case~(\ref{EFF1}) is still the leading contribution
from the zero modes (constant modes on the coset). In this way of counting,
the non-zero mode contributions are Gaussian and subleading in the large 
volume limit, making (\ref{EFF1}) universal \cite{GASSER,SMILGA}.
In the microscopic regime, Leutwyler and Smilga 
\cite{SMILGA} further observed that when (\ref{EFF1}) is converted 
through $Z_n (M)= \int d\theta/2\pi e^{-i\theta n} 
Z(\theta, M )$
to the n-vacuum state, the 
terms of order $(VM\Sigma )^k$ in the Taylor expansion of $Z_n$ are in 
one-to-one correspondence with inverse moments of powers of the eigenvalues 
of the Euclidean and massless Dirac operator, 
$iD\!\!\!/[A]\phi_k =\lambda_k[A]\phi$,
in the external gluon field $A$. A typical example is the first
moment~\cite{SMILGA}
\be
\left< \sum_i' \frac 1{\lambda_i^2[A]}\right>_n = 
\frac {V^2\Sigma^2}{4(|n|+N_f)}
\label{sum1}
\ee
where the averaging is over the quenched gluon field, excluding the zero
modes (primed sum), in the large volume $V$ limit~\footnote{The
ultraviolet singularities of~(\ref{sum1}) are subleading in
$V$~\cite{SMILGA}. This justifies the decoupling assumption between soft
and hard modes for the inverse moments in the microscopic limit.}.  In
the n-state, the number of Dirac eigenvalues $\lambda_k$ near zero is
decreased with increasing windings $|n|$ and flavors $N_f$, due to the
accumulation of zero modes (exclusion principle).  More importantly,
(\ref{sum1}) implies that the spectrum near zero involves a huge number
of eigenvalues $\lambda_k\sim 1/V$, in comparison to $1/\sqrt[4]{V}$ in
free space. The onset of spontaneous chiral symmetry breaking is
followed by a large accumulation of eigenvalues in the microscopic
region near zero virtuality, a point first made by Banks and
Casher~\cite{BANKSCASHER}.

\subsection{Universal Spectral Fluctuations in Vacuum}

Random matrix models share~(\ref{EFF1}) with QCD at tree level 
(macroscopic regime) and to leading order (microscopic limit), so
they also share these moments. This is also true for the effective 
models announced earlier, such as the instanton model~\cite{ALKOFER} 
and the Nambu-Jona-Lasinio model~\cite{NAMBU}. Unlike other models,
however, random matrix models provide a {\it simple} access to a
chiral fermionic spectrum where the properties following from the
spontaneous symmetry breaking are properly encoded at zero virtuality. 
(The same features can be checked to hold for the Nambu-Jona-Lasinio
model or the instanton model on an Euclidean hypertorus, provided that
the ultraviolet contributions in the former are removed.).

The set of all moments introduced by Leutwyler and Smilga follows  
readily from a {\it mean} spectral density $\overline{\nu} (\lambda)$
which follows from~(\ref{EFF1}) as suggested in~\cite{SMILGA}. Indeed,
for $N_f=1$ (see footnote) and in a fixed n-vacuum~\cite{SMILGA}
\be
\frac 1V \overline{\nu}_n (\lambda ) = \sum_{k=0}^{\infty} \delta
	(\lambda V \pm \frac{\xi_{n,k}}{\Sigma})
\label{MICRO}
\ee
where $\xi_{k,n}$ are the positive zeros of $J_n (\xi_{n,k}) =0$ with 
$\xi_{n,k}\sim \pi (k +n/2-1/4)$. For $V\rightarrow\infty$ and 
$x=\lambda V$ fixed,  (\ref{MICRO})
reproduces ${\it all}$ of the sum rules discussed in~\cite{SMILGA}. This
limit is referred to as microscopic, and is best taken using random matrix 
models~\cite{SHURYAKVER,BESSEL}. For continuum four dimensional QCD in the 
0-vacuum state, the microscopic distribution is given by Bessel 
functions~\cite{BESSEL},
\be
\nu_s (x) = \frac {x\Sigma^2}2
\left( J_{N_f}^2 (\Sigma x) -J_{N_f+1} (\Sigma x) J_{N_f-1} (\Sigma x)\right)
\label{MICROS}
\ee
with a node at zero in agreement with (\ref{sum1}) and higher moments.
The larger $N_f$, the wider the node near zero. Eq.~(\ref{MICROS})
captures the onset of the spontaneous breaking of chiral symmetry
near zero virtuality as suggested by~(\ref{MICRO}). This result was 
extended to the n-vacuum state in \cite{VERBESSEL}.

Although the microscopic distribution was constructed using a Chiral 
Gaussian Unitary Ensemble (ChGUE), it was postulated in \cite{BESSEL}
to be {\it universal}, and therefore shared by QCD itself. There, it
was also argued that for sufficiently random gauge configurations, 
the distribution
of eigenvalues is a priori constrained by the central limit theorem in the 
microscopic regime. This assumption, however, is not needed. Any chiral
model is bound by chiral power counting to reproduce~(\ref{EFF1}) and
hence~(\ref{MICRO}). This includes chiral random matrix models with an
arbitrary polynomial weight consistent with unitary invariance 
(unbroken chiral symmetry), and positivity. Non-Gaussian weights bring about
corrections to~(\ref{EFF1}-\ref{MICRO}) that vanish as $V\rightarrow\infty$.
Hence~(\ref{EFF1}) uniquely constrains both the diagonal and
off-diagonal parts of the $n$-point eigenvalue-correlators 
in the microscopic limit, forcing them to be essentially unique (universal). 
This point has been explicitly proven in~\cite{NISHIGAKI}. 

In fact chiral symmetry through~(\ref{EFF1}) also constrains the fluctuations
around~(\ref{MICRO}) to be generic, since the finite size
corrections in $1/{V}$ are likely to be subleading compared to the bulk
distribution, in the regime where 
$|\langle\bar{q}q\rangle|\sim M V\Sigma^2$. These 
fluctuations are best analyzed by unfolding the spectra \cite{MEHTA}, that 
is reexpressing the distribution of states around the mean,
say~(\ref{MICRO}), and calculating the pertinent
spectral statistics. We believe that this is what transpires from the 
detailed calculations in \cite{UNFOLVER} for noncontinuum QCD. Our
observations do not extend to the regime where 
$|\langle\bar{q}q\rangle|\sim \Sigma$ except if chiral
power counting is only enforced at tree level, or if the volume
corrections are accidentally small.

{}From our analysis, it is clear that~(\ref{EFF1}) is a result that holds for
quenched and unquenched spectra of the Dirac operator, provided that the
associated ground state breaks spontaneously chiral symmetry in the 
macroscopic limit. This is the case for QCD, since chiral symmetry is
believed to be 
spontaneously broken even 
in the quenched case. The effect of the unquenching, is to provide for chiral 
loops etc. , but these effects bring about corrections to~(\ref{EFF1}) that 
vanish exponentially in the large volume limit. This observation confirms a 
detailed analysis of the effects of unquenching on the microscopic spectral 
density~\cite{JUREK}. Changes in the microscopic spectral density are induced 
in the transition regime $M V {\Sigma}\sim 1$. This is expected from chiral 
power counting since the loop effects are no longer suppressed in this
regime, $e^{-\sqrt[4]{V} M}\sim e^{-1/\sqrt[3]{V}}\sim 1$. In this 
case~(\ref{EFF1}) receives contributions from loops in the infrared,
and is no longer unique. This is the regime where the dynamical content of
the theory is tested, and where lattice and continuum physics are likely to
depart.

We note, that all our arguments apply to continuum QCD. On the lattice,
the issue of chiral symmetry is blurred by the fermion doubling problem
and the nature of the Lorentz structure on a discrete mesh (staggered 
formulations). 
Nevertheless, some of the above concepts may be extended to this case as well.
However, we expect various microscopic universalities to set in between
the strong and weak coupling limits. Indeed, for staggered quarks in strong
coupling, the exact (but bare) chiral symmetry is $U(1)\times U(1)$, and is 
only expected to turn to $U(N_f)\times U(N_f)$ in weak coupling. The strong
to weak coupling limit is followed by a reordering of the quark eigenmodes
near zero virtuality. This point can be analytically investigated along the 
lines
put forward by Gross and Witten~\cite{GROSS}, and checked numerically using
present lattice formulations. The true continuum limit is only recovered 
in weak coupling with full Lorentz and chiral symmetry restored.

Finally, it is worth mentioning at this stage that in odd-dimensional
spaces the  
issue of chiral symmetry is more subtle. Nevertheless, the spontaneous 
breakdown of a global form of chiral symmetry is possible, and an analogue
effective action to~(\ref{EFF1}) for QCD in three dimensions
has been worked out, together with its microscopic distribution~\cite{QCD3}. 
This is an important point, since the nature of matrix models in QCD
encompass the instanton issue from which they were originally concocted 
(see below). There are no instantons in odd-dimensional spaces.

\subsection{Universal Spectral Fluctuations in Matter}

Under changes in the external parameters such as $N_f$, 
$N_c$, $M$, $\theta$, temperature, quark chemical potential, QCD may undergo 
phase changes. If we were to assume that some  of these phase changes obey 
the general lore of universality, then at the critical points the transition 
is dictated by the symmetry of the order parameter (again the nature of 
the coset space) and a number of relevant operators as originally discussed 
by Pisarski and Wilczek~\cite{PISARSKI}. While the relevant operators 
characterize the structure of the `effective potential' (radial variable),
the nature of the coset in the weak field limit dictates the universality
of the spectral fluctuations of the QCD Dirac operator in matter. 
For instance, a phase change induced 
by the temperature $T$ would yield around the critical point to ($\beta=1/T$)
\be
Z ( \theta, M; T) \sim \int\,\, [d\Phi ] \,\, e^{-\beta V_3\,\,{\cal L}_3 
(T, \theta, M, \Phi )}
\label{EFF111}
\ee
where ${\cal L}_3$ involves polynomial operators
\footnote{The possibility of phase changes induced by non-polynomial
operators,
$e.g.$ see~\cite{KOCIC}, will not be discussed here.}
\be 
{\cal L}_3 ( T , \theta, M, \Phi )&=& {\rm Tr} (e^{i\theta/N_f} M \Phi + 
\Phi^{\dagger} M^{\dagger}e^{-i\theta/N_f} )
+ g_0(T) \,\,{\rm Tr} (\Phi^{\dagger}\Phi )
 \nonumber\\
&+&  g_1 (T) {\rm Tr} (\Phi^{\dagger}\Phi )^2 + g_2 (T)
({\rm Tr} (\Phi^{\dagger}\Phi ))^2 \nonumber\\
&+& g_3 (T) ({\rm det}\Phi +{\rm det}\Phi^{\dagger}) + {\cal O} (\Phi^6 )\,.
\label{EFF11}
\ee
Here $\Phi$ is a complex $N_f\times N_f$ matrix, $V_3$ a dimensionless 
parameter commensurate with the three volume, 
and ${\cal O}$ are contributions from higher dimensional operators
as well as non-zero modes.
$g_0 (T)$ is negative below and zero at the critical temperature. 
The standard and nonstandard chiral random matrix models formulated
so far are consistent with (\ref{EFF11}) near the critical point 
\cite{USNJL}. 

We observe that near the critical temperature, the relevant operators
determine the value of $|\Phi |\sim \Sigma (T)$ at the saddle point
in the radial direction. In the weak field limit, this value is
$M$ independent, and as a result (\ref{EFF111}-\ref{EFF11}) reduce 
to (\ref{EFF1}) with the substitution $\Sigma\rightarrow \Sigma (T)$,
$V\rightarrow \beta \,V_3$, and $SU(N_f)\rightarrow U(N_f)$. 
In the microscopic regime $\beta\,V_3 M\Sigma (T)\ll 1$, 
the vacuum sum rules are unchanged provided that the
concept of an n-state remains valid (the $\theta$ integration merely bringing
an $M$-independent overall normalization for $g_3\neq 0$).  This is 
the case for the chiral random matrix models discussed in \cite{USNJL}. 
These arguments are in agreement with detailed calculations 
in the case $N_f=1$ and $g_3=0$ \cite{TEMPMICRO}. Following on our arguments
in section 3.1.2, we would also expect the fluctuations in the infrared part
of the bulk of the spectrum to be universal, and follow from chiral random
matrix models. This can also be checked by studying the spectral statistics
of the unfolded spectra at finite temperature. 

Clearly, our arguments apply 
to other phase changes provided that the analysis is maintained within the
spontaneously broken phase (approach from below), for which the concept of
an order parameter such as $\Phi$ makes sense
\footnote{For phase transitions in terms of $N_f$ and $N_c$, we should think 
about interpolating in these parameters. Also at finite chemical potential
the microscopic limit has to be defined with care.}.
In this sense there is a {\it dichotomy} in universality: 
the universal spectral oscillations of the QCD
operator in the infrared are tied to the universality of phase transitions
at the critical point. What is amazing, is the fact that the spectral 
oscillations remain unaltered below the phase transition,
thanks to the universality of phase changes.

\section{APPLICATIONS of BLUE'S FUNCTIONS}

Having gone over the universal aspects of spectral oscillations in the
microscopic limit of effective QCD, we now return to some aspects of the
macroscopic limit, where random matrix models are regarded as effective 
models beyond the contribution of order $M$ and infinite volume. In this 
regime, the $1/N$ approximation is an interesting way of organizing the 
expansion (modulo exceptional points), where subtleties may arise from 
different limiting procedures (large volume, small masses, quenched, etc.).

For effective QCD, chiral random matrix models offer a useful alternative
to conventional power counting as is the case for the U(1) and CP problem, 
and provide a simple framework for discussing phase changes driven by 
conventional and non-conventional universality arguments. Although they do
not follow the general lore of conventional chiral power counting, they are 
still useful
in the infrared for a qualitative description of bulk observables for which 
soft and hard modes decouple (see below).
For models of quantum mechanics with or without dissipation it 
is a useful mean for implementing mean-field-like approximations. We now 
proceed to illustrate these points.

\subsection{Chiral Random Matrix Models}

Chiral random matrix models follow naturally from the 0-dimensional
reduction of chiral four-fermi models in even dimensions.
Specifically~\cite{USNJL}
\be
{\cal L}_0 = q_L^{\dagger} iM e^{i\theta/N_f} q_R + 
             q_R^{\dagger} iM e^{-i\theta/N_f} q_L
+ \frac 1{{\Sigma V}} q_L^{\dagger}q_L q_R^{\dagger}q_R 
\label{FERMI}
\ee
where ${\rm dim}\,\, q_{R,L}= N_fn_{+,-}$ characterizes the dimension of the 
Grassmanian space, and $n_+ + n_- = V$. Other `multi-quark' interactions are
possible provided that they transform as singlets $(1,1)$ under 
$SU(N_f)\times SU(N_f)$\footnote{For $N_f\geq 2$ with the U(1) broken by
determinantal-type interactions. The present arguments hold for $N_f=1$ as 
well~\protect\cite{USNJL}.}. Chirality
is the only remnant of the even dimensionality of the original space.
For odd dimensions, we refer to~\cite{QCD3}.
In the macroscopic regime, these models owe their essential physics to the 
instanton liquid model (from which they were originally inspired)
\cite{BESSEL,USPAST,SHURYAKVER,SIMONOV}, and in the 
microscopic limit to the uniqueness of~(\ref{EFF1}), following
from~(\ref{FERMI}) by bosonization $W = q_Lq_R^{\dagger}$.

Indeed, from present cooled lattice simulations, it appears that the bulk
features of a number of hadronic expectation values and correlations can be
well-described by an ensemble of random topological structures 
(singular instantons) where the quark and antiquark zero modes play an
important 
(perhaps) dominant role. The hopping between the zero modes in this complex
environment can be {\it assumed} to be random with a typical strength set by 
the chiral condensate in the vacuum. This is a non-controllable approximation,
and in general causes most of the results in the macroscopic limit to depend
on the nature of the random weight. If we were  to assume that the QCD vacuum
is sufficiently random, then there may exist an infrared Gaussian fixed point 
that will cause the weight to truncate to a Gaussian, in the macroscopic
limit~\footnote{The existence of an infrared Gaussian fixed point in the
macroscopic limit may be addressed using the renormalization group
method discussed in~\cite{RENORM}.}. Since instanton models are fair
effective models 
to the QCD vacuum and its low-lying excitations, we may expect chiral random 
matrix models to be a fair (albeit cruder) effective model for bulk 
observables with weak ultraviolet sensitivity. Their advantages being 
their {solubility} and {clarity} in the thermodynamical limit.

The bosonized version of the four-fermi model~(\ref{FERMI}) 
yields $W =q_Rq_L^{\dagger}$, that is
\eqn
Z(\theta , M )=\sum_{n_{\pm}}\int[dW] & &e^{-N/\sigma{\rm Tr} {\cal V}
(WW^{\dagger})}
 \cdot e^{-\frac{\chi^2}{2\chi^{\star}V}}
 \cdot e^{-\frac{\kappa^2}{2\kappa^{\star}V}}\cdot\nonumber\\
&&\cdot\prod_{N_f} {\rm det}
\left( \begin{array}{cc}
        m_fe^{i\theta/N_f} & iW \\ iW^{\dagger} & m_fe^{-i\theta/N_f}
        \end{array} \right) \,.
\label{QCDdef}
\eqnx
Here $W$ is a complex random asymmetric $n_- \times n_+$ matrix. The
two diagonal blocks have implicit size $n_{\mp} \times n_{\mp}$. The
weight for the random potential is assumed Gaussian for simplicity.
The widths of the Gaussians: $\sigma$, $\chi^*$, $\kappa^*$ are related to 
the quark condensate, the topological susceptibility and the compressibility 
of the quenched ensemble, respectively. They have been discussed 
in~\cite{US}. 
Here $2N=n_++n_-$, $\kappa \pm \chi = 2(n_{\pm}-\langle n_{\pm}\rangle)$, 
$N_f$ is the
number of flavors, and $V$ a dimensionless parameter commensurate with the
original four volume. For most cases we will assume $2N/V=1$ in the 
thermodynamical limit unless stated otherwise.

The model~(\ref{QCDdef}) has been studied
extensively~\cite{BESSEL,USPAST,SHURYAKVER,ALKOFER,SIMONOV,US,USU1} and
this is why we have chosen to analyze it in these lectures. It is
minimal, in the sense that it follows from the instanton model to the
QCD vacuum in the zero-mode sector and infinite wavelength
limit~\cite{USPAST}. To account for the presence of near zero modes,
along with the topological (exact) zero modes, a simple variation
of~(\ref{QCDdef}) is just~\cite{USNJL}
\footnote{We are restricting the discussion of~\cite{USNJL} to the
vacuum with a $\theta$ angle. The inclusion of matter in~(\ref{QCDdef2})
only affects the near zero modes, ensuring exact zero modes throughout,
as expected from QCD.}
\be
Z (\theta, M) = \sum_{n_{\pm}} \,\,\xi (n_+, n_-) \,\,
\int \, d{\bf R} \,e^{-\frac{\Sigma}2 \,{\rm Tr} ({\bf R}{\bf 
R}^{\dagger} )} \, {\rm det}{\bf Q}
\label{QCDdef2}
\ee
with $\xi (n_+, n_-)$ a pertinent measure in the quenched limit, 
and ${\bf Q} = {\bf D} + {\bf R}$. The random 
matrix ${\bf R}$ has rectangular entries  
\be
{\bf R} =
\left( \begin{array}{cccc}
        0 & {\bf A} & 0 & {\bf \Gamma}_R^{\dagger} \\
        {\bf A}^{\dagger} & 0 & {\bf \Gamma}_L^{\dagger} & 0\\
        0 & {\bf \Gamma}_L & 0 & {\bf \alpha} \\
        {\bf \Gamma}_R & 0 &{\bf \alpha}^{\dagger} & 0\end{array}\right)
\label{ST1}
\ee
and the sparse and deterministic matrix ${\bf D}$ has square entries,
\be
{\bf D} = 
\left( \begin{array}{cccc}
        iMe^{i\theta/N_f} & 0& 0 & 0 \\
        0& iMe^{-i\theta/N_f} & 0 & 0\\
        0&0&iMe^{i\theta/N_f}&0\\
        0&0&0&iMe^{-i\theta/N_f}\end{array}\right) \,.
\label{ST2}
\ee
Each of the four rows and columns in the above
matrices~(\ref{ST1}-\ref{ST2}) have dimensions $N_f N$, $N_f N$,
$N_fn_+$, $N_fn_-$, respectively. ${\bf A}$ characterizes the hopping
between {\it near} zero modes, $\alpha$ the hopping between the zero
modes, and ${\Gamma}$ the cross-hopping between zero and near zero
modes. For ${\Gamma }=0$ the near zero modes decouple from the zero
modes in this model. This non-standard model is perhaps more in line
with the recent QCD studies on the
lattice~\cite{URBANA}. Since~(\ref{QCDdef2}) share the same~(\ref{EFF1})
with QCD in the vacuum, it fulfills the same microscopic sum rules, and
hence the same Bessel oscillations~\cite{BESSEL}.  It is amusing to note
that a similar class of random matrices (ring matrices) was actually
discussed by Br\'{e}zin, Hikami and Zee in some generic models of
disorder~\cite{RING}, where Bessel oscillations identical to the ones
following from~(\ref{QCDdef}) were derived in the microscopic regime,
for the case of symmetric matrices with $n_+=n_-=N$ and $N_f=1$.

\subsubsection{Decoupling and Spectral Densities}

We note that both (\ref{QCDdef}) and (\ref{QCDdef2}) are chiral 
in the sense that the argument of the determinant anticommutes 
$\gamma_5\equiv {\rm diag} ({\bf 1}_{n_+}, -{\bf 1}_{n_-}) $, 
in the massless case. For simplicity we return to (\ref{QCDdef})
and restrict for the moment to the case $N_f=1$. For $\theta=0$ and
fixed $\kappa$, the spectral properties of (\ref{QCDdef}) follow from
\be
G(z)=\frac{1}{2N} \left< \frac{1}{z-
\left( \begin{array}{cc}
        m & iW \\ iW^{\dagger} & m
        \end{array} \right)
} \right> \,.
\label{greenrec}
\ee
Using (\ref{CUTX}), we have ($m\rightarrow 0$)
\be
\left< \bar{q} q \right>  = -\frac 1N \sum_n \frac m{m^2+\lambda_n^2} =
 -\frac {\pi}{N} \,\nu(\lambda =0 )
\label{BANKS}
\ee
where the density of states $\nu(\lambda )$ relates to the discontinuity
of $G(z)$ through (\ref{CUTX}) with $m=0$.
For the quark condensate to be non-zero, the average number of 
quark states near zero virtuality has to grow with the size $N$. 
(\ref{BANKS}) means that the level spacing near zero is of 
order $1/N$ in comparison to $1/\sqrt[4]{N}$ in free space.
Chiral symmetry brings about huge correlations near zero virtuality.
The relation (\ref{BANKS}) was originally established by Banks and 
Casher~\cite{BANKSCASHER}. It is also the first of the 
infinite sum rules discussed recently by Leutwyler and Smilga~\cite{SMILGA}.

Since the spacing near zero is of order $1/N$ it follows that 
in the microscopic limit $N\lambda\sim 1$, the density of states is solely
given by the soft modes, as the hard modes (essentially free) are separated
to infinity. In this limit the spectrum near zero is blown up
to cover the whole energy scale, causing the hard modes to
decouple by hand. The random matrix assumptions of keeping
only the soft modes become legitimate. In the macroscopic limit, 
the hard and soft modes couple, and the decoupling assumption
is in general not justified. However, for a number of infrared
sensitive parameters this coupling is weak. For instance, although
(\ref{BANKS}) diverges quadratically, the divergences are readily 
seen to be of the form ${\bf C}_1 m + {\bf C}_2 m^3$ by performing
a double subtraction. Hence, they do not affect explicitly 
the value of the quark condensate in the chiral limit, as expected
for an order parameter. They only affect it implicitly through the occurrence
of anomalous dimensions, which is a mild logarithm dependence on the
renormalization scale. These arguments carry on to matter, 
thanks to the non-renormalization theorems \cite{LANDSMAN}.

The averaging over the rectangular matrices in (\ref{greenrec})
could be performed either by using diagrammatic methods~\cite{USU1}, 
or by using the multiplication law for Blue's 
functions by rewriting the rectangular matrices as square random matrices
multiplied with projectors (see next section). In both cases, $G(z)$ takes
the generic form
\be
G=g {\bf 1} + g_5 \gamma_5 
\ee
for fixed $n_{\pm}$ with  ${\bf 1} \equiv {\rm diag} ({\bf 1}_{n_+},
{\bf 1}_{n_-}) $. Specifically
\be
{\rm Tr }\,G&=&g+xg_5 = \frac{1}{2}(z-\sqrt{z^2-4+4x^2/z^2})|_{z\equiv im}
\nonumber \\
{\rm Tr }\,G\gamma_5&=& xg+g_5 =-\frac{x}{z}|_{z\equiv im}
\label{both}
\ee
for the chirality even and odd resolvents with
asymmetry  $x=(n_+ -n_-)/2N$. The corresponding spectral densities 
are
\be
\nu_+(\lambda, x)&=&|x|\delta(\lambda) +\frac{1}{2\pi |\lambda|}
\sqrt{(\lambda^2-\lambda^2_-)(\lambda^2-\lambda^2_+)} \nonumber \\
\nu_-(\lambda, x)&=&x\delta(\lambda)
\label{spectralboth}
\ee
with $\lambda^2_{\pm}= 2\pm2\sqrt{1-x^2}$.
For $x=0$ we recover Wigner's semi-circle. In particular, $\nu_+ (0)\neq 0$
in agreement with (\ref{BANKS}) and the general lore of spontaneous breaking
of chiral symmetry in the infinite volume limit. Note that $\nu_- (0)$
measures directly the difference between the left-handed and right-handed
quark zero modes. It follows from the rectangularity of the
chiral random matrices, and is a schematic version  of the Atiyah-Singer
index theorem~\cite{ATIYAHSINGER}.

\subsubsection{Screening of the Topological Charge}

The quenched version of (\ref{QCDdef}) shows that the topological charge is 
fixed by the width of the Gaussian, that is $\chi \sim \chi_*\sim m^0$ and
finite in the chiral limit. In the presence of quarks, the Gaussian 
distribution gets correlated with the fluctuations in the asymmetry of the 
matrices. As a result, the topological charge $\chi\sim m$ and vanishes
in the 
chiral limit. What causes this is the occurrence of terms of order $x/m\sim 
1/Nm$ in the large $N$ analysis which makes the limits $N\rightarrow \infty$
and $m\rightarrow 0$ non-commutative. 

 To analyze this consider the vacuum partition function (\ref{QCDdef}) for 
fixed asymmetry $x=\chi/2N$, equal quark masses and zero $\theta$. For
$z=im$, we have
\be
\partial_z \log Z_{\chi} =-2iNN_f\, {\rm Tr} G
\ee
where ${\rm Tr} G$ follows from (\ref{both}). $Z_{\chi}$ follows through the
integration law  (\ref{blueint}) or (\ref{trick}) for Blue's functions. 
However, for the screening of the topological charge we only need to retain
the terms of order $x^2\sim 1/N$ in $G(z)$, since the Gaussian forces
$\chi\sim \sqrt{N}$. With this in mind, we have 
$G(z)\approx -x^2/(z^2\sqrt{z^2-4})$, so that integration of Green's
(Blue's)
function yields
\be
\log Z_{\chi}=\log Z_{\chi=0}+\frac{N_f}{2N}
	\frac{1}{m(\sqrt{m^2+4}+m)}\chi^2 +{\cal O}(\chi^3) \,.
\label{part2}
\ee
The integration constant was fixed from the asymptotic condition on 
$Z_{\chi}$.
Inserting $Z_{\chi}$ into ($V=N$)
\be
Z (0 , M) =\int d\chi \,\,\,e^{\frac{-\chi^2}{2\chi^* N}}\,\,\,Z_{\chi}
\label{XPART}
\ee
yields the following substitution
\be
1/\chi^* \rightarrow 1/\chi^* +
\sum_{i=1}^{N_f}\frac{1}{m_i(\sqrt{m_i^2+4}+m_i)} \equiv 1/\chi^* 
+\sum_{i=1}^{N_f} \gamma_i 
\label{screening}
\ee
after reinstating the flavor dependence.
For $N_f=0$ (quenched) the topological susceptibility is just 
given by $\chi^*\sim m^0$, while for $N_f\neq 0$ it is screened,
$\chi\sim m\ll 1$.
The fluctuations of order $x^2$ induced by quark matrices of different
sizes are crucial for this screening. This mechanism is 
generic~\cite{GENERIC},
and holds for more realistic models of the QCD vacuum such as the instanton 
vacuum model~\cite{INSTSCREENING}.

\subsubsection{U(1) Problem}

The non-vanishing of the unquenched topological charge is usually enough to 
show that QCD solves its U(1) problem.  
This is also true in the present chiral
random matrix models. Indeed, at finite $\theta$, (\ref{QCDdef}) allows us to
define a simple `Ward-identity' between the topological susceptibility
$\tilde{\chi}$, the pseudoscalar susceptibility $\chi_{ ps}$ and the
quark condensate \cite{USU1}. For $\theta=0$ and equal quark masses, this 
identity is just
\be
\tilde{\chi}=-\frac{2m}{N_f^2}i {\rm Tr G}-\frac{m^2}{N_f^2}\chi_{ps}
\ee
in total analogy with the QCD Ward identity
\be
i\chi_{top}=-\frac{im}{N^2_f}\langle\bar{\psi}\psi\rangle 
+\frac{m^2}{N^2_f}\int d^4x \langle T^* \bar{\psi}i\gamma_5 \psi(x)
\bar{\psi}i\gamma_5 \psi(0)\rangle
\label{wardqcd}
\ee
where the pseudoscalar susceptibility is defined by the integral 
in (\ref{wardqcd}). The resolution of the U(1) problem in QCD stems from
the fact that the unquenched topological susceptibility
 $\chi_{ top}\neq 0$, for
otherwise (\ref{wardqcd}) implies that the powers of $m$ matches only if
the pseudoscalar susceptibility develops an infrared sensitivity in the
form of $1/m$, much like the pion susceptibility. In the chiral random matrix 
model, the analogue of (\ref{wardqcd}) is
\be
\chi_{ps} \equiv - \frac{1}{N} \langle{\rm tr} {\bf Q}^{-1}\gamma_5 
{\bf Q}^{-1}\gamma_5\rangle
+\frac{1}{N} \langle{\rm tr} {\bf Q}^{-1}\gamma_5{\rm tr} 
	{\bf Q}^{-1}\gamma_5\rangle_{\rm conn.} 
\label{nont1}
\ee
where ${\bf Q}$ denotes the argument of the determinant
in~({\ref{QCDdef}). From the preceding discussion, it follows that
\be
  \chi_{ps} \sim (\sum_i^{N_f}\frac{1}{m_i})^2\cdot \frac{1}{1/\chi^*
	+\sum_i^{N_f}\gamma_i} -4\sum_i^{N_f}\gamma_i
\label{nontrivial}
\ee 
which is {\it finite} in the chiral limit.  For illustration, we
show in Fig.~\ref{f.u1} the behavior of the various terms in
(\ref{nontrivial}) following from ensemble averaging large samples of
random matrices of different sizes $N$ versus the analytical results
(solid curve) in the large $N$ limit.

The remarkable cancelation of the the $1/m^2$ terms in (\ref{nontrivial})
between the trace-term and the trace-trace term, implies that the disconnected
term is as important as the connected one in this channel, contrary to 
conventional wisdom. 
 The reason is the strong infrared sensitivity in this channel as
emphasized above. Crucial to our argument is the issue of the large $N$
limit or the thermodynamical limit since $V\sim N$ in our analysis. The
above issues are important, and should be retained in a lattice analysis
of full QCD, as they teach us valuable lessons on the subtle interplay
between fluctuations in finite size lattices and the onset of screening.

\begin{figure}
\centerline{\epsfxsize=12cm \epsfbox{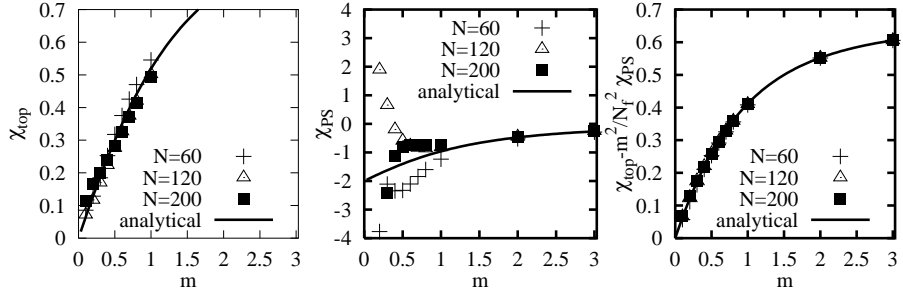}}
\caption{Topological (left), pseudoscalar (middle) susceptibilities
and anomalous Ward identity (right) for three equal flavors.
The numerical simulations were carried for for fixed
$2N=n_++n_-=60,120,200$. Solid lines are analytical results.
}
\label{f.u1}
\end{figure}

\subsubsection{$\theta$ Vacua}

For $\theta\neq 0$ the determinant in (\ref{QCDdef}) is complex. Simulations
in QCD with complex fermionic determinants are still elusive, owing to the
occurrence of random phases. In simplified models, we refer to the recent
numerical studies in \cite{SCHIERHOLTZ}. For this problem, chiral random 
matrix models 
may be effective in addressing some of the underlying issues, especially
those related to the thermodynamical limit. Having
said this, we note that (\ref{QCDdef}) is periodic of period $2\pi$ in 
$\theta$, provided that the sums $n_{\pm}$ span over n-vacua with 
$n$ even {\it and} odd. We also note, that in general the matrices sampled in 
(\ref{QCDdef}) are finite in size, thereby making the issue of the $\theta$
dependence of the vacuum state subtle in the thermodynamical limit 
\cite{USTHETA}. In a nutshell: the precision of the calculation and the
thermodynamical limit are intertwined, and erroneous results may emerge
if due care is not taken \cite{USTHETA}. The large $N$ analysis we will 
present below has been found to be consistent with the above observations
\cite{USTHETA}.

With this in mind, and to investigate the behavior of the vacuum
energy versus $\theta$ we now proceed to estimate ${\rm ln} Z (\theta , 
M)/N$. If we were to assume that the range of resummation over 
$\chi=n_+-n_-$ is infinite, for $N= n_++n_-$ peaked with a small but
nonzero spread, then the Gaussian integration in (\ref{QCDdef}) can be 
readily performed. Standard bozonization~\cite{USTHETA} gives
\be
{\rm ln} Z (\theta, M) = N\,{\rm Tr\, ln} \left( -\frac{|z+P|^2}{|P|^2}\right)
+\frac {\chi_*V}{8}\left({\rm Tr\, ln}\frac {\overline{z} +P^{\dagger}}
{z+P}\right)^2
\label{AP1}
\ee
where $z={\rm diag}\,\,\, m_je^{i\theta/N_f}$, and $P$ is a complex 
$N_f\times N_f$  matrix expressed in terms of the $m_j$'s and $\theta$
through the saddle point equations (see below). In the vacuum, $P={\rm
diag}\,\,\, 
p_je^{i(\theta/N_f-\phi_j)}$, so that the unsubtracted free energy associated 
to (\ref{AP1}) reads
\eqn
F (\theta, M)  =&& \sum_{j=1}^{N_f} \left( {\rm ln}\, p_j^2 
-{\rm ln}(p_j^2 + 2m_j p_j {\rm cos}\, \phi_j + m_j^2 )\right) \nonumber\\
&& +\frac {\chi_* V}{2N} \left(\theta + \frac i2
\sum_{j=1}^{N_f} {\rm ln}
\left( \frac{m_j+p_je^{i\phi_j}}{m_j+p_je^{-i\phi_j}} \right)\right)^2 \,.
\label{AP2}
\eqnx

{}From (\ref{AP2}) it follows immediately that if the quarks masses are
large, that is $m_j\gg p_j\sim 1$ at the saddle points, then the 
$\theta$ dependence
of the unsubtracted free energy is just that of the quenched limit, that is
$F\sim \chi_*\theta^2/2$. For small masses $m_j\ll\chi_*\sim p_j\sim 1$
(the scale of spontaneous symmetry breaking), then (\ref{AP2}) simplifies to
\be
F (\theta, M) = &&\sum_{j=1}^{N_f} \left( p_j^2 -{\rm ln}\, 
 p_j^2 -2\frac {m_j}{p_j} {\rm cos}\, \phi_j \right) \nonumber\\
&&+\frac {\chi_* V}{2N} \left(\theta -\sum_{j=1}^{N_f} \phi_j
-\sum_{j=1}^{N_f} \frac {m_j}{p_j} {\rm sin} \phi_j \right)^2
+ {\cal O} (m^2) \label{AP3} \,.
\ee
The saddle point solution in the $p$'s decouples and gives
$p_j=1+{\cal O} (m)$. The saddle points in the $\phi$'s give 
\be
&&\theta=\sum_{j=1}^{N_f} \phi_j +{\cal O} (m)\nonumber\\
&&m_1{\rm sin}\, \phi_1= ...=m_{N_f} {\rm sin}\, \phi_{N_f}
\label{AP4}
\ee
in agreement with results derived by Witten~\cite{WITTEN} using large 
$N_c$ (number of colors) arguments. That they are reproduced
by~(\ref{QCDdef}) 
is reassuring. This also means that they should also be satisfied in the
instanton model as well as some variants of the Nambu-Jona-Lasinio model with 
proper $U(1)$ breaking \cite{ALKOFERZAHED}.

For $N_f=2$ these equations can be readily solved to give
\be
{\rm sin}\phi_{1,2} =\pm \frac {m_{2,1} {\rm sin}\, \theta}
{\sqrt{m_1^2 +m_2^2 + 2m_1m_2 {\rm cos}\,\theta}}\,.
\label{AP5}
\ee
As a result $Z (\theta , M)$ is a smooth $2\pi$ periodic function except 
at $\theta=\pi$ and $m_1=m_2$ where both the numerator and denominator
vanishes in (\ref{AP5}). In this case, the free energy 
is again $2\pi$ periodic but with a cusp at $\theta=\pi$. The cusp
reflects on a first order transition caused by the coexistence of two
CP violating solutions. As a result CP is spontaneously broken at $\theta=\pi$
in agreement with earlier arguments \cite{WITTEN,DASHEN,NYUTS,CREUTZ}. 

For $N_f=3$, the explicit solutions to (\ref{AP4}) are in general involved
analytically. However, at $\theta=\pi$ the analysis simplifies. There are two 
classes of solutions. In general, a unique solution that is CP
symmetric, and a doubly 
degenerate solution for $m_1m_2> m_3 |m_1-m_2|$ for which CP is spontaneously 
broken, again in agreement with a number of analyses
\cite{WITTEN,DASHEN,CREUTZ}.

\subsubsection{Finite Temperature}

The formalism of Blue's functions is very well suited to investigate
the random matrix models of the QCD Dirac operator for finite
temperatures and/or chemical potential. 
Indeed, the modifications in  the medium corresponds 
to replacement of  the argument of the determinant (modulo trivial
factor of $i$)
(\ref{QCDdef}) by 
\eq
\arr{0}{{\bf \Omega}+i\mu}{{\bf \Omega} +i\mu}{0}+\arr{0}{W}{W^{\dagger}}{0}
\label{repla}
\eqx
where $\mu$ is the chemical potential for ``quarks'' and 
${\bf \Omega}=\omega_n {\bf 1}_n \otimes {\bf 1}_N
$. Here $\omega_n = (2n+1)\pi T$  are all fermionic Matsubara frequencies. 
The form of the replacement in (\ref{repla}) is consistent with the effective
models of QCD and the Dirac structure~\refnote{\cite{USNJL}} near the critical
temperature. For simplicity, we restrict ourselves to the lowest pair of 
Matsubara  frequencies $\pm\pi T$ and the chiral limit. This corresponds 
to the model proposed in~\refnote{\cite{JV}}.

To investigate chiral symmetry breaking in this
model we should calculate the Green's function and, through
the Banks-Casher relation, analyze the chiral condensate near the 
critical point. Therefore, we have to ``add'' 
 the deterministic (D) and random (R) 
``Hamiltonians'', 
\be
\arr{0}{\pi T}{\pi T}{0} + \arr{0}{W}{W^{\dagger}}{0} \,.
\label{diracT}
\ee
In the present form, the Blue's function
technique does not incorporate the chiral block structure of the
matrices, but for the deterministic matrices considered here it
works correctly~\refnote{\cite{USBLUE}}. The Blue's function 
for~(\ref{diracT}) satisfies
\eq
B(z)-z=B_D(z)
\eqx
where $B_D$ is the Blue's function of the deterministic
piece. By definition we have
\eq
z-G(z)=B_D(G(z)) \,.
\eqx
Now if we evaluate the Green's function of the {\em deterministic}
piece on both sides of this equality we will get Pastur equation~\cite{PASTUR}
\eq
G_D(z-G(z))=G_D(B_D(G(z))) \equiv G(z) \,.
\label{eq-past}
\eqx 
The  explicit form of the deterministic Green's function is 
$G_D(z)=z/(z^2-\pi^2 T^2)$, since the deterministic eigenvalues are $\pm
 \pi T$, so the final result reads~\footnote{This cubic equation is shared 
by chiral and non-chiral
random matrix models alike~\protect\cite{PASTUR}. The chiral structure
shows up only in $1/N$.}
\eq
\label{cardano}
G^3-2zG^2+(z^2-\pi^2 T^2+1)G-z=0 \,.
\eqx
The `critical temperature' for chiral symmetry restoration follows from the
the condition $B'(G)=0$. Using (\ref{cardano}) this locates the end-points
of the quark spectrum at
\be
\pm A_1&=&\frac{1}{\sqrt{8}\pi T}
\frac{ (4\pi^2T^2-1+\sqrt{8\pi^2T^2+1})^{\frac{3}{2}}}{\sqrt{8\pi^2T^2
+1}-1 }
\nonumber \\
\pm A_2&=&\frac{1}{\sqrt{8}\pi T}
\frac{(4\pi^2T^2-1-\sqrt{8\pi^2T^2+1})^{\frac{3}{2}}}{\sqrt{8\pi^2T^2
+1}+1} \,.
\ee
The condition $A_2=0$ corresponds  to the situation
when the two-arc support of the spectrum $[-A_1, -A_2] \cup [A_2,A_1]$
evolves into a one-arc at $[-A_1,+A_1]$. These structural changes in
the spectrum of quark eigenvalues are shown in Fig.~\ref{f.spodnie}.
The critical temperature is $T_*=1/\pi$, in units where the width of the
random Gaussian distribution is set to 1.
The break up near zero virtuality $(\lambda\sim 0)$ may be generic 
of a phase change in the system, provided that the {\it decoupling}
between soft ($S$) and hard ($H$) modes take place in the Dirac 
spectrum~\cite{USNJL}. This is the case for order parameters as
we discussed above.

\begin{figure}[t]
\centerline{\epsfxsize=12cm \epsfbox{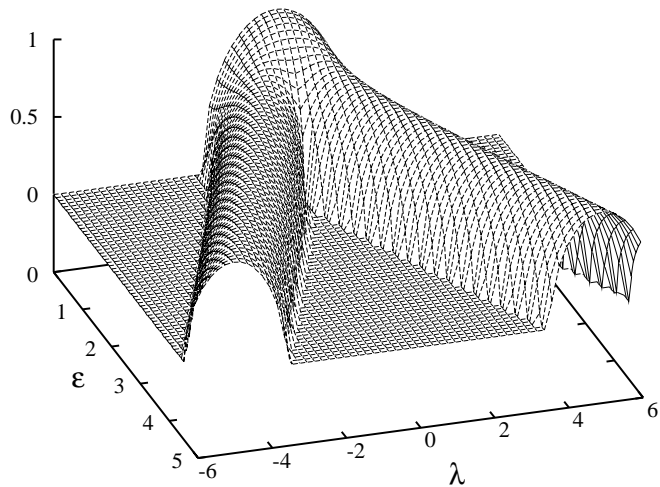}}
\caption{Spectral density $\nu(\lambda)$ as a
 function of the deterministic parameter
$\epsilon=\pi T$.}
\label{f.spodnie}
\end{figure}

The set of all critical exponents could be easily 
inferred from solving the algebraic equation (\ref{cardano}) 
and calculating the free energy of the system by integrating 
Blue's functions. The result is
\eq
\label{free}
F=G^2+\log \f{z-G}{G} \,.
\eqx
All critical exponents are of the mean-field-type in agreement with 
conventional universality arguments~\refnote{\cite{USNJL}}. It is not
difficult to construct random matrix models with non-mean field critical
exponents. Using the results in \cite{USNJL} one can easily see that the
effective potential at the critical point is all what matters, much like
the effective potential (\ref{EFF1}) at zero temperature as discussed in
section 3.1. A non-mean field
effective potential can be unraveled in the form of a chiral random matrix
model by using the {\it inverse} procedure. An example of a model with first 
order transition was given in \cite{USNJL}.

How realistic are all these statements? This can be seen by comparing
some of the present results to lattice simulations near the critical
point. Recent simulations by the Columbia group \cite{CHRIST}, have 
revealed interesting insights on the distribution of eigenvalues using
the semi-quenched condensate
\be
\langle\overline{\zeta}\zeta\rangle =2m_{\zeta}
\int_0^{\infty} \, d\lambda\, \frac{\nu (\lambda )}{\lambda^2 + m_{\zeta}^2}
\label{SEMI}
\ee
where the distribution of eigenvalues is for $N_f=2$ with sea quark masses
$m_s\neq m_{\zeta}$. 
Although at finite $m_{\zeta}$ there is a sensitivity of (\ref{SEMI}) to the
hard modes (see above), the latter is suppressed for small $m_{\zeta}$.
The behavior of (\ref{SEMI}) for staggered fermions on a $16^3\times 4$ 
lattice is shown in Fig.~\ref{fig-lat} (left) versus the valence quark
mass $m_{\zeta}$ for equal but fixed sea quark masses $m_sa=0.01$ (6 MeV). 
In the chiral random matrix model, the eigenvalue distribution $\nu (\lambda)$
follows from (\ref{eq-past}) through the substitution $\pi^2 
T^2\rightarrow\pi^2T^2 + m_s^2$ \cite{USLATTICE}. The result for (\ref{SEMI})
is shown in Fig.~\ref{fig-lat} (right) for $m_s=0.1$ (10 MeV). We have 
identified the lattice $\beta$ with $T$ through : $\pi 
(T-T_c)=(\beta-\beta_c)$ and set $\beta_c=5.275$. There is overall 
qualitative agreement over several decades of the valence quark mass
$m_{\zeta}$.

\begin{figure}[htbp]
\centerline{%
\epsfysize=55mm \epsfbox{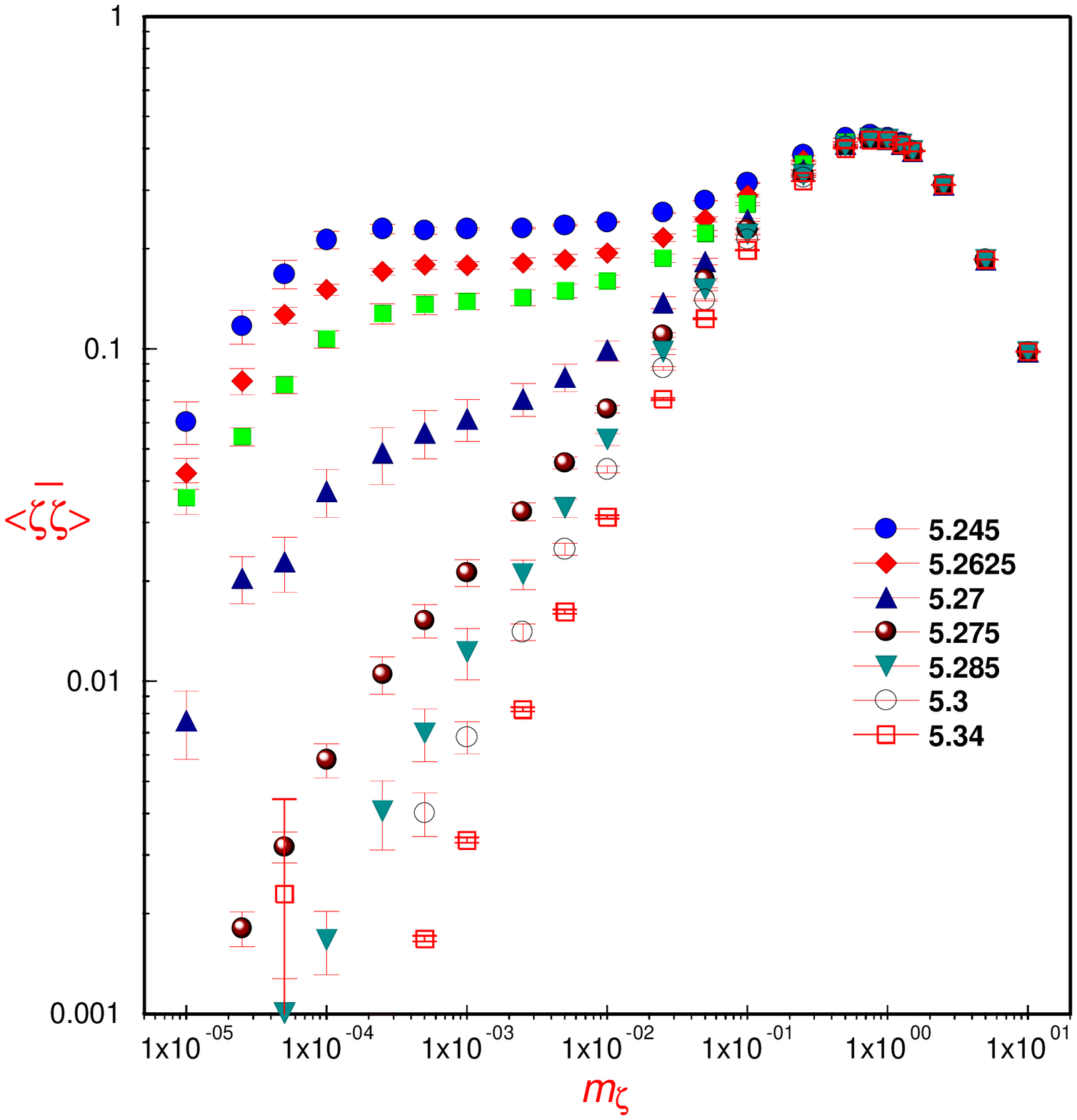} \hfill
\epsfysize=57mm \epsfbox{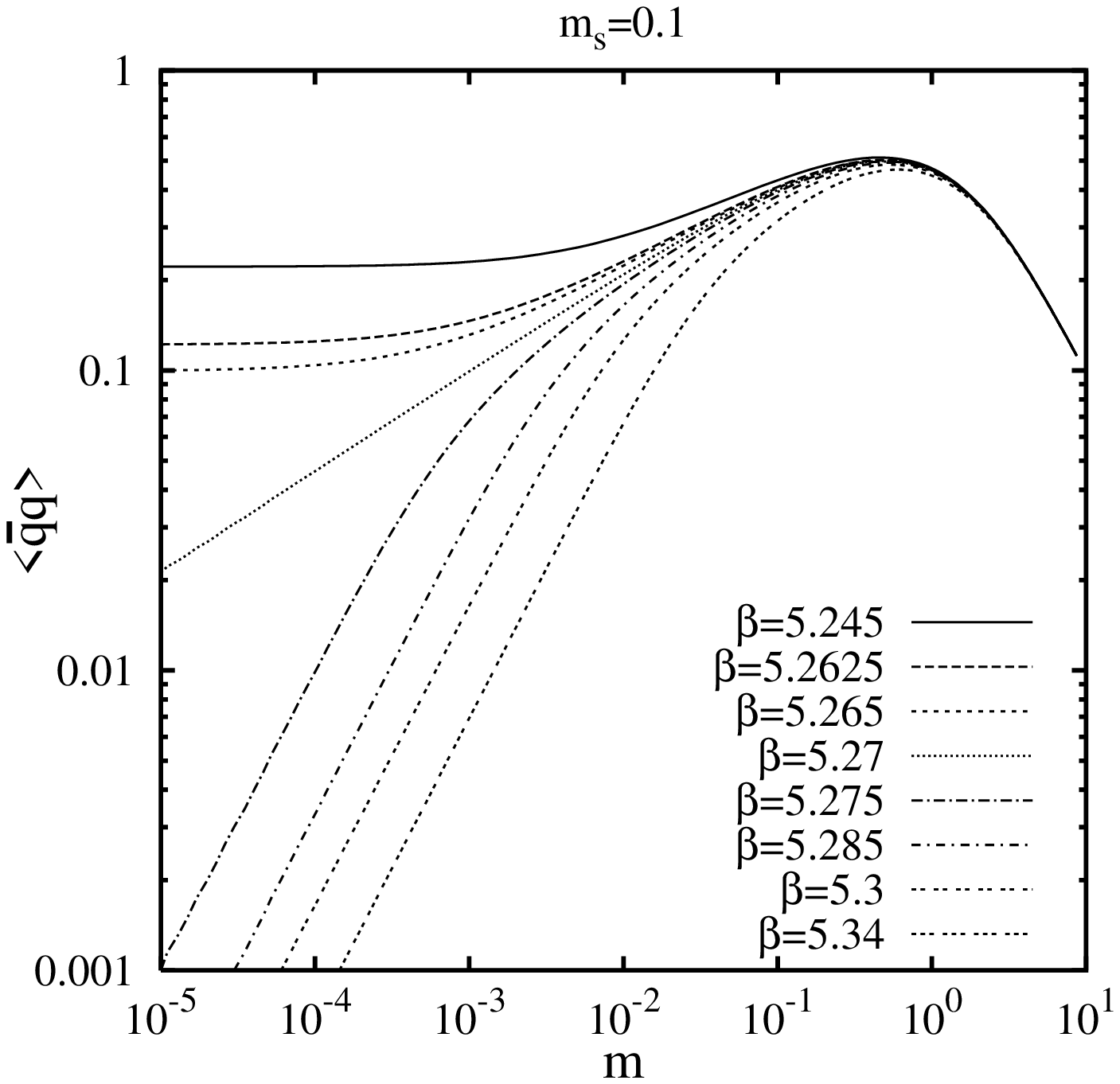} \hfill}
\caption{Semi-quenched condensate versus the valence quark mass $m_{\zeta}$
for two-flavor QCD \protect\cite{CHRIST} with a sea mass $m_sa =0.01$
(left). The result on the right is from a chiral random matrix model. 
See text.}
\label{fig-lat}
\end{figure}

\subsubsection{Finite Chemical Potential}

The case of finite chemical potential can be analyzed in a similar way, using 
instead
\eq
\label{diracmu}
\arr{0}{i\mu}{i\mu}{0}+\arr{0}{W^\dagger}{W}{0} \,.
\label{RANDMU}
\eqx
The deterministic part is now nonhermitian and we
have two distinct regions (at least in the quenched case) on the complex
plane. The
Green's function in the {\it holomorphic region} outside the blob of
eigenvalues is given by (\ref{cardano}) with the substitution $\pi^2
T^2 \ra -\mu^2$:
\eq
\label{cardanomu}
G^3-2zG^2+(z^2+\mu^2+1)G-z=0 \,.
\eqx
In order to find the boundary of the nonholomorphic
region we may exploit the method of conformal mapping
via the Blue's functions, and transform the cuts of the 
$T\neq 0$ case into the boundary (see Fig.~\ref{f.map}) 
by the transformation
\eq
z \ra iw= z-2G(z)
\label{MAPPINGX}
\eqx
where $G(z)$ is the appropriate branch of the cubic equation
(\ref{cardano}) with the {\it formal} identification $\pi^2T^2=+\mu^2$. 
The extra factor of $i$ corresponds to the rotated Hamiltonian (here the
deterministic as opposed to the random part is antihermitian).
\begin{figure}
\centerline{\epsfysize=60mm \epsfbox{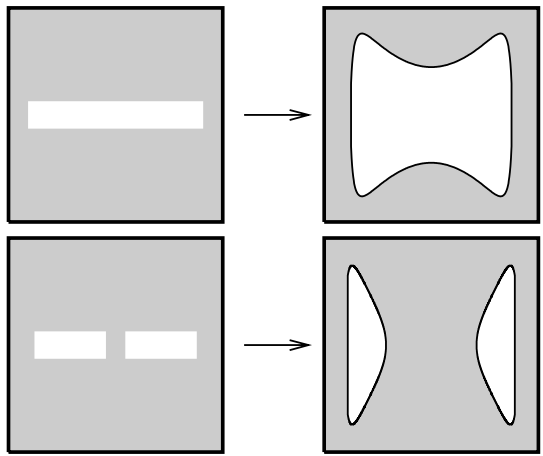}}
\caption{Conformal mapping from temperature to chemical potential.
The one- and two-arc cuts (left) are mapped onto islands (right).
The shaded region is the holomorphic support of the mapping.}
\label{f.map}
\end{figure}

The result of this mapping is in total agreement with a result 
derived originally by Stephanov~\refnote{\cite{STEPHANOV}} using
a variant of the replica technique for the identical problem. The 
domains encircled by the mapping (\ref{MAPPINGX}) defines the
support of the complex eigenvalues of (\ref{RANDMU}). In this 
region the resolvent $G(z)$ is non-analytic. The analytical 
resolvent following from (\ref{cardanomu}) defines the correct
branch in the holomorphic (shaded region), but fails on the
boundary where the $1/N$ expansion breaks down. The failure is
caused by the accumulation of eigenvalues in the plane, and
signals large fluctuations in the phase of (\ref{RANDMU}) as
well as its correlations~\refnote{\cite{USMUX}}. Similar effects are 
observed in quenched lattice calculations~\cite{MULAT}.  

This effect is an artifact of the quenching (suppression of the fermion 
determinant) in the calculation of the resolvent which is a {\it singular}
operator \cite{DIAG}. This may be overcome by unquenching at the price of
much larger precision in the numerics, due to the presence of phase 
oscillations \cite{USUNPUBLISHED}, or analyzing {\it non-singular} operators
such as the partition function, which are self-quenched in the large $N$ limit
\cite{DIAG,USNJL,BOOK}. High precision calculations for the
{\it unquenched} partition function $Z_N (z=0, m)$ with $N_f=1$ are shown in 
Fig.~\ref{f.zhol} (dotted line). The solid lines refer to the two
holomorphic solutions of~(\ref{cardanomu}). The long arrows indicate the
points where 
the islands of non-analyticity in the quenched resolvent cross the real axis.
The short arrow is the position of the cusp resulting from 
the intersection of the two holomorphic solutions
to~(\protect\ref{cardanomu}).
\begin{figure}
\centerline{\epsfysize=37mm \epsfbox{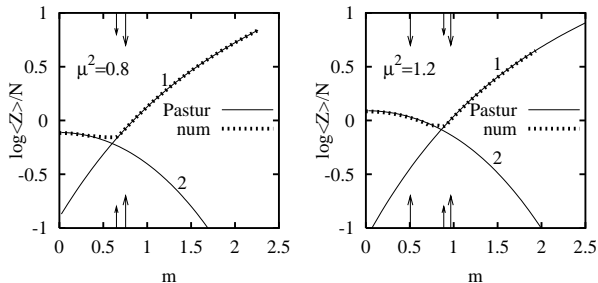}}
\caption{Unquenched free energy for $N_f=1$ at $\mu^2=0.8$ (left)
and $\mu^2=1.2$ (right). The dots are high precision numerical
results, while the solid lines follow from the solutions
to~(\protect\ref{cardanomu}). See text.}
\label{f.zhol}
\end{figure}

The phase structure of~(\ref{RANDMU}) 
directly follows from the analytical properties of the partition function 
$Z_N (z, \mu )$ for general complex $z=im$ with $N_f\neq 0$ and in the
$1/N$ approximation as also noticed in \cite{STEPHANOV}. In our case, this 
can be readily done using the integration law for 
Blue's functions and the holomorphic solutions to (\ref{cardanomu}) 
\cite{DIAG}. The outcome is simple: $Z_N$ develops cusps in the 
$z$- or $\mu$-plane that reflect on the two-branches of (\ref{cardanomu}),
with end-points conditioned by $B' (G)=0$, in agreement with the results
in \cite{STEPHANOV}. The curves of zeroes obtained in
this way are also in agreement with the (high-precision) numerics 
of \cite{HJV}. The location of all zeroes follow from the analytical 
condition (\ref{cuspline}), here 
\be
{\rm Re} \left[ 
G^2_i(z)-G^2_j(z) +\log \frac{z-G_i(z)}{z-G_j(z)} 
+\log \frac{G_j(z)}{G_i(z)}\right] =0
\label{ZER}
\ee
where the labels $i,j$ refer to the different branches of the solutions
(\ref{cardanomu}). The density of zeroes follows accordingly from 
(\ref{densdel}).

Using similar techniques one could also calculate the effect of fluctuations
($1/N$ corrections to the one-point Green's function), 
with the result~\cite{DIAG}
\be
Z_N(z,\mu)=e^{NE_0}\cdot \left(\left\{D^{-2}[(D^2+\mu^2)^2
-(z-G)^4]\right\}^{-\frac{1}{2}} + {\cal O}\left(\frac{1}{N}\right)
\right) \,.
\label{Zloop}
\ee
Both $E_0$ and $D$ depends functionally on $G$ to this order.
Hence, the information carried by the
Blue's functions condition the analytical behavior of (\ref{Zloop}). 
In particular, the location of the zeros of the curly bracket coincide
with the points at which the $1/N$ expansion breaks down. These are
exceptional points that may signal the onset of new scaling regimes 
for a new class of microscopic universalities in the context of 
non-hermitian random matrix models~\cite{DIAG,NATO}.
\begin{figure}
\centerline{\epsfxsize=12cm \epsfbox{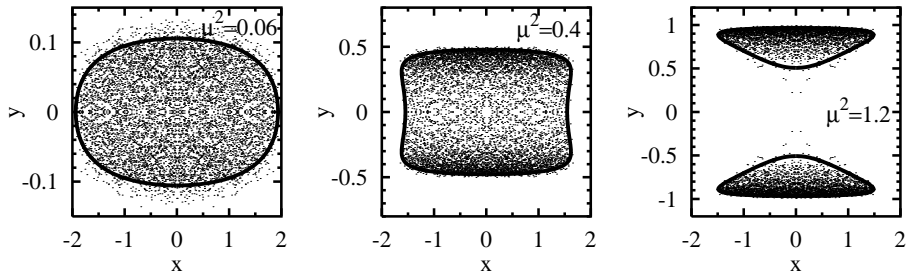}}
\caption{Sample distribution of eigenvalues for different $\mu^2$.
The solid lines follow from the conformal mapping.}
\label{f.chem}
\end{figure}

\subsubsection{Phase-Diagram}

For general finite $T$ and finite $\mu$ the discussion about the
thermodynamics of chiral random matrix models is very much model
dependent, much like the thermodynamics in the Nambu-Jona-Lasinio model
or the instanton liquid model.  It is simply that of `constituent
quarks', with one quark and one anti-quark state (two-level)
\cite{USNJL,BOOK}. In the absence of matter, chiral symmetry is
spontaneously broken by filling the `negative-energy' state. In the
presence of matter, the symmetry is restored by depleting the
`negative-energy' state and filling the `positive-energy' state with
equal weight (finite T), or when the quark chemical potential reaches
the `positive-energy' state (finite $\mu$). In the first case the
transition is second order, while in the first case it is first
order. Diquarks effects in the present chiral random matrix models are
down by $1/N$, and hence they decouple in the large $N$ limit.  They are
inessential for the thermodynamics in these models.  Although unwelcome
for $N_c \geq 3$ QCD because of color confinement (zero triality), they
are dominant for $N_c=2$ QCD where they play the role massless
baryons.~\footnote{Alternative chiral random matrix models with diquark
states to order $N^0$ are straightforward to write down. They will be
discussed elsewhere.}

Having said this, we now consider (\ref{repla}) with infinitely many
Matsubara modes. This is essential when departing from the critical
points, or in taking the temperature to zero for fixed $\mu$
\cite{USNJL}.  In the chiral limit, the quark condensate
$|\langle\overline{q} q\rangle| = P_*$ (spectral density $\nu (0)$),
follows from the saddle point equation to the partition function
associated to (\ref{repla}), that is~\cite{USNJL,BOOK}
\be
2\Sigma P_*=1-\frac{1}{1+e^{(|P_* +m| -\mu)/T}}
-\frac{1}{1+e^{(|P_* +m| +\mu)/T}} \,.
\label{trans}
\ee
The solution to (\ref{trans}) for $m=0$
is shown in Fig.~\ref{njl.map}. For $\mu>1/6\Sigma \ln{(2+\sqrt{3})}$, the 
transition is
first order and disappears at $\mu_*=1/2\Sigma$. At $\mu=0$, the transition 
is second order (mean-field) and sets in at $T_*=1/4\Sigma$. The
mismatch with
the $T_*=1/\pi$ quoted in section 3.1.6 is due to: a different normalization 
for $\Sigma$ and an infinite number of Matsubara modes. 
The behavior shown in Fig.~\ref{njl.map} is overall consistent with results 
in four dimensions using constituent quark models, $e.g.$ \cite{NAMBU}. We 
refer to \cite{USNJL,BOOK} for further details on some of these and other
issues in relation to phase changes. We note that the result for $\mu=0$ 
was also reproduced by \cite{WETTIGWEINDEN} using a similar random matrix
model. The effects of the inclusion of more Matsubara modes was also
investigated in~\cite{HIP}.

\begin{figure}
\centerline{\epsfysize=60mm \epsfbox{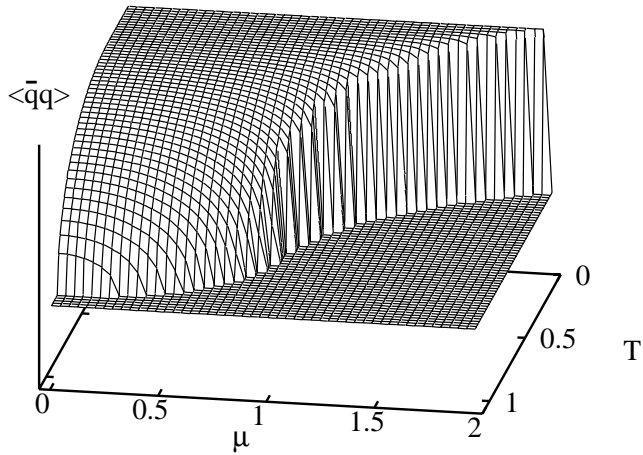}}
\caption{$|\langle\bar{q} q\rangle|$ from~(\protect\ref{trans})
with $\Sigma=1/4$ and $m=0$.}
\label{njl.map}
\end{figure}

\subsection{Strongly Nonhermitian Ensembles}

The theory of open quantum systems plays a key role in many areas of
physics and chemistry~\refnote{\cite{EWA1to6}}.  Microscopic treatment
of dissipative evolution involves hermitian Hamiltonians. However, the
use of {\it non-hermitian} Hamiltonians has been found to provide a
convenient way to give a {\it reduced} description of the system. In
this case, the hermitian part of the Hamiltonian refers to the free
undamped dynamics whereas the nonhermitian remainder describes the
damping imposed on the system by some external source.  In this section,
we will show a few applications of the Blue's functions for dissipative
systems, where the effective Hamiltonian is obtained by the standard
Wigner-Weisskopf~\cite{WEIS} reduction by partition. Reduction by
partition follows from dividing the whole Hilbert space into two
subspaces, and then the Hamiltonian is integrated over one of them. In
this way one reduces the eigenvalue problem of high dimension to a lower
one, and the resulting nonhermicity of the {\it effective} Hamiltonian
comes from the ``leakage of probability'' due to the truncated vectors
of the original Hilbert space.

\subsubsection{Open Chaotic Scattering}
We start from non hermitian random matrix model introduced by Mahaux
and Weidenm\"{u}ller~\cite{MAH} for resonance scattering. 
In
brief, a quantum system composed  of $(N-M)$ closed and $M$ open channels
can be described by the effective scattering Hamiltonian
\be
H= H_R + H_{\Gamma} \equiv H_R -ig \Gamma \qquad,\qquad \Gamma= AA^T
\label{defgam}
\ee
which is $N\!\times\!N$ dimensional. 
$A$ is an asymmetric $N\!\times\!M$ random matrix, $H_R$ is random
Gaussian
Hamiltonian and $g$ an overall
coupling. Unitarity enforces the form of $\Gamma$ used in~(\ref{defgam}).
The matrix elements
$A_k^a$ characterize the transition between the $(N-M)$-internal channels
and $M$-external channels, and are assumed to be independent of the
scattering energy. For $M=0$, the Hamiltonian is real
and the spectrum is bound. For $M\neq 0$, the Hamiltonian is complex, with
all states acquiring a width. 
In what follows, we will solve the system by ``decomposition'' 
into  subsystems and consecutively apply the methods presented
above for Blue's functions.

First consider the case of an $N\!\times\!N$ real symmetric
product matrix $\Gamma^S_{kl} = A^a_k A^a_l$. 
Note that  the Green's function (for even random potentials) 
for the square of a square matrix is  related to the
Green's function for the unsquared one via 
$G(z^2=w)=G(z)/z$, due to the identity
\be
G(w)=\left<{\rm Tr} \frac{1}{w-A^2} \right>=
\frac{1}{2z}\left<{\rm Tr} \frac{1}{z-A} \right> 
+\frac{1}{2z} \left<{\rm Tr} \frac{1}{z+A} \right> \,.
\label{trickone}
\ee
{}From (\ref{trickone}) and (\ref{gauss}), we infer
the resolvent  for the square of the Gaussian 
\be
G_{\!A^{\!2}} = \frac{1}{2} \left( 1 - \sqrt{1-\frac{4}{z}} \right) \,.
\label{square}
\ee

Second, let us observe, that the  case of rectangular $N\times M$ matrices
with $N, M\rightarrow \infty$ but $m=M/N$ fixed, 
follows by truncation using the projector~\cite{VOIREC}
\be
P={\rm diag}(\underbrace{1,\ldots,1}_{M},\underbrace{0,\dots,0}_{N-M})
\label{projector}
\ee
that is $\Gamma = \Gamma^S P$.
We recognize  the problem of ``multiplication'' of the random 
matrix $\Gamma^S$ by the deterministic projector P, and therefore could
use the multiplication law for the Blue's functions, hence the 
S-transform.
The construction of the  resolvent for the projector is straightforward 
\be
G_P(z) \equiv \frac{1}{N} {\rm Tr}\frac{1}{z-P} 
= m \frac{1}{z-1} + (1-m) \frac{1}{z}\,.
\ee
Using the relations for ``multiplication law'' from the previous part 
we obtain the S-transform 
\be
S_P(z) = \frac{1+z}{m+z}\ .
\ee
Similarly, the S-transform for the Green's function 
for the square of the Gaussian (\ref{square}) is 
\be
S_{\!A^{\!2}} = \frac{1}{1+z}\ .
\ee
The product reads
\be
S = S_{\!A^{\!2}} \cdot  S_P =\frac{1}{m+z} 
\ee
so inverting  the order of reasoning  we get for the resolvent
\be
G(z)=\frac{1-m}{2 z} + \frac{1}{2} \left[
	1 \pm \sqrt{\left( \frac{1-m}{z}-1 \right)^2 -\frac{4m}{z}} \right] \,.
\label{resrec}
\ee
Note that the $1/z$ terms (first term and a term under the square root)
represent $(N-M)$ zero modes of the matrix $\Gamma$.  

Now we could ``add''  the Hamiltonians $H_R$ and $-ig\Gamma$, 
by constructing the generalized Blue's functions.
Let us first  write down the hermitian Blue's functions for $H_R$
and $g\Gamma$.
The first Blue's function, coming from inverting (\ref{gauss}) is 
\be
B_1(z)=z+\frac{1}{z} \,.
\label{bluegauss}
\ee 
The second one is the functional inverse of (\ref{resrec}), therefore
\be
B_2(z)= \frac{m g}{1-gz}+\frac{1}{z} \,.
\label{twoblue}
\ee
Then the  {\it generalized} Blue's functions read
\be 
\BB_1(\ZZ)&=&\ZZ+\frac{1}{\ZZ} \nonumber \\
\BB_2(\ZZ)&=& m(1-\hat{g}\ZZ)^{-1}\hat{g} +\frac{1}{\ZZ}
\label{genblue1}
\ee
where the coupling $\hat{g}$ is a matrix $\hat{g}={\rm diag} (-ig, ig)$.
The signs in the matrix coupling reflects the fact, that the 
complex conjugate ``copy'' interacts with opposite coupling constant.
Then, the generalized addition law for the Blue's functions leads
to the final matrix equation for the Green's function
\be
\ZZ= m(1-\hat{g}\GG)^{-1}\hat{g} +\GG+\frac{1}{\GG} \,.
\label{finalhaake}
\ee
Nontrivial (nonholomorphic) solution to the matrix equation
(\ref{finalhaake}) provides full information on the eigenvalue
distribution. The
condition $G_{q\bar{q}}=0$ defines the boundaries of the eigenvalue
distribution, and the divergence of $G_{qq}$ (Gauss law) gives the desired
spectral density of eigenvalues on the islands. Instead of presenting
the explicit formulae, we show the figures with few sample solutions
(figure \ref{f.haake}).

\begin{figure}
\centerline{\epsfxsize=12cm \epsfbox{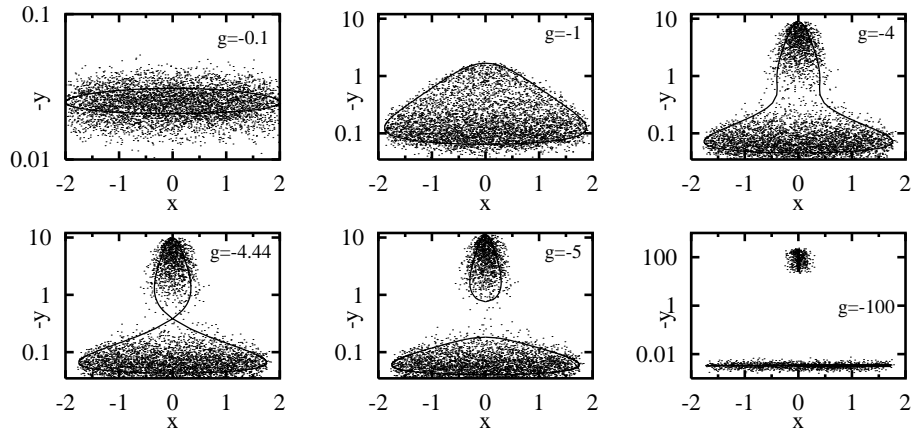}}
\caption{Comparison of the numerically generated distribution of
eigenvalues for few sample couplings in the model~(\protect\ref{defgam})
 with the
analytical results. The solid lines represent the boundaries following 
from the condition $G_{q\bar{q}}=0$.}
\label{f.haake}
\end{figure}
 
The results obtained above were obtained also by other 
methods~\cite{OTHERHAAKE}, so we do not
need to dwell on their physical content. However, we encourage
the reader to compare the present derivation with the derivations
of the same result based on the replica or supersymmetric methods (see
e.g.~\cite{SOMMERSREV,EFETOV}). 

\subsubsection{Bridging via Dissipation} 

We conclude this brief overview by another example, where analytical results
were obtained by direct application of the generalized Blue's functions.
This is the model of an effective two-level system, coupled to a noise 
reservoir of the random type~\refnote{\cite{EWA}}. Such systems could 
e.g. serve as model systems for electron or proton transfers in condensed 
phase reactions in the field of chemical physics. The deterministic part 
of the Hamiltonian is given by an effective two-level system with N 
electrons, 
building up two close energy states with half of the energies equal to 
$\epsilon$ and other half equal to $-\epsilon$. For the dissipative part, we 
take $H_{\Gamma}$ from (\ref{defgam}), therefore the effective Hamiltonian 
reads 
\be
H=H_D - ig AA^{\dagger} \,.
\label{ewaham}
\ee
The deterministic Green's function for two levels is 
\be
G_D=\frac{1}{2}(\frac{1}{z-\epsilon} +\frac{1}{z+\epsilon})
\label{detgreen}
\ee
therefore the relevant generalized Blue's function is given by the matrix 
equation 
\be
\ZZ=\frac{1}{2}\left[(\BB_D(\ZZ)-\epsilon)^{-1}+
(\BB_D(\ZZ)+\epsilon)^{-1}    \right] \,.
\label{detblue}
\ee
The generalized Blue's function for the noise term is just $\BB_2$ in 
(\ref{genblue1}). The generalized addition law for the Blue's functions
leads at once~\cite{EWA}
\be
\GG=\frac{1}{2}\left[(\ZZ-m(1-\hat{g}\GG)^{-1}\hat{g} -\epsilon)^{-1}
+(\ZZ-m(1-\hat{g}\GG)^{-1}\hat{g} +\epsilon)^{-1}
\right] \,.
\label{blueewa}
\ee
As in the previous example, one easily infers the spectral density of
the system 
(\ref{ewaham}) from the nonholomorphic solution to the equation. Again, we 
present a figure instead of the explicit formulae (see figure \ref{f.ewa}).

\begin{figure}
\centerline{\epsfxsize=12cm \epsfbox{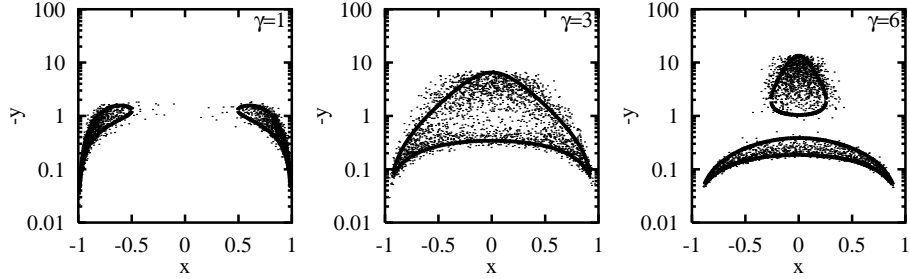}}
\caption{Comparison of the numerically generated distribution of
eigenvalues for few sample parameters of the model (\protect\ref{ewaham}).
The solid lines represent the boundaries following from 
the analytical condition $G_{q\bar{q}}=0$.}
\label{f.ewa}
\end{figure}

We would like to note two ``phase transitions'' visible easily from the change 
of character of the envelopes. Here the envelope is represented by the solid 
curve, 
obeying the analytical solution to $G_{q\bar{q}}=0$, here  given by the 8th 
order polynomial in $x$ and $y$ ( with $z=x+iy$). The first phase transition 
(figure \ref{f.ewa}(left)$\rightarrow$ \ref{f.ewa}(middle)) corresponds 
to the emergence of the noise-induced ``bridging states''. The second
phase transition, 
present at large coupling constant (figure \ref{f.ewa}(middle)
$\rightarrow$ \ref{f.ewa}(right)) corresponds to the ``collectivization 
of the spectra''\cite{SOKZEL}. The upper island represents  direct 
resonances, the lower 
island represents the (long-lived) coherent state, caused by the 
explicit presence of zero modes in the dissipative part of the Hamiltonian.

\subsection{Weakly Nonhermitian Ensembles}

As a final application of the Blue's function techniques we will
consider the model of Hatano and Nelson~\refnote{\cite{HATANONELSON}} (HN)
for nonhermitian localization, which has recently attracted much attention
\cite{HNUS,HNOTHER,UPDATENH}. This model is a non-hermitian version of the
Anderson model, hence quantum mechanical in nature. Nevertheless, the 
method of 
Blue's functions will turn out to be very appropriate for its treatment, 
since it is equivalent to the Coherent Phase Approximation (CPA) widely used
in the treatment of hermitian models with site disorder~\cite{CPA}. 

In their original analysis, Hatano and Nelson \cite{HATANONELSON} have
shown that the depinning of the flux lines from columnar defects in 
superconductors in $d+1$-dimensions  may be mapped  onto the world-lines 
of bosons on a $d$-dimensional hyperlattice. The localization by defects 
was approximated by an on-site randomness on the hyperlattice, and the
depinning induced by the transverse magnetic field was found to cause
a {\em directed} hopping of the bosons on the hyperlattice, resulting
in a nonhermitian quantum Hamiltonian.

\begin{figure}
\centerline{\epsfxsize=12cm \epsfbox{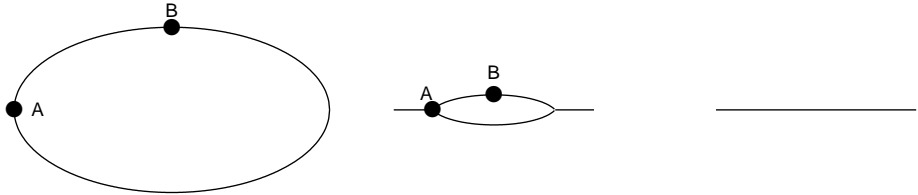}}
\caption{Eigenvalue distribution in the Hatano-Nelson model for 
large (left), intermediate (middle) and small (right)  $h$. The two dots refer
to the location of the forking (A) and the maximal imaginary eigenvalue (B).}
\label{f.spectrum}
\end{figure}

The HN model (for $d=1$) is given by the following Hamiltonian
\eq
\label{HNhamilt}
H=H_0+\VV=\sum_{A=1}^N  \f{t}{2}\left( e^h c^\dagger_{A+1}c_A+e^{-h}
c^\dagger_{A-1} c_A\right)+V_A c^\dagger_A c_A
\eqx
where the $c^\dagger_A$ is a boson creation operator at site $A$. The
lattice spacing is set to 1. The hopping strength is $te^{\pm h}$, where
$h$ is the strength of the ``transverse magnetic
field''~\refnote{\cite{HATANONELSON}} in units of the flux
quantum. Diagonal randomness is represented by random variables $V_A$,
originally taken from a uniform distribution in the interval
$[-\Delta,\Delta]$. For $h\neq 0$, the eigenvalues are complex.  As
stated earlier, this is a nonhermitian analogue of the Anderson model in
$d=1$ dimension. The eigenstates of this Hamiltonian with an eigenvalue
$\lm$ such that ${\rm Im} \lm\neq 0$ are delocalized. For sufficiently
large $h>h_c^{(2)}$ all states are delocalized and form a closed curve
that looks like a deformed ellipse (see figure
\ref{f.spectrum}(left)). When $h$ is reduced below $h_c^{(2)}$, the
eigenvalues at the edges snap to the real axis and get localized (see
figure \ref{f.spectrum}(middle)). Reducing $h$ further below $h_c^{(1)}$
causes the eigenvalues to fill in the whole interval on the real axis
(see figure \ref{f.spectrum}(right)).  Our aim here is to understand
these effects quantitatively, using the method of Blue's functions.

The relevant quantity to calculate is the Green's function
\eq
G(z)=\f{1}{N}\cor{\tr\f{1}{z-H}} \,.
\eqx
Its holomorphic part is related to the derivative of $\det(z-H)$ 
outside (large $z$). A close look at (\ref{HNhamilt}) shows that
before averaging over $\VV$, 
\eq
\det(z-H)\sim \left( \f{t}{2} \right)^N \cdot 
\left[ e^{Nh}+e^{-Nh}\right]+f(z,N,\VV)
\eqx
where the function $f$ does not depend on $h$ and has the leading
order behavior 
\eq
f(z,N)=e^{N f_{out}(z,\VV)}\cdot (1+{\rm corrections})
\eqx
where the corrections are polynomial.
For large $N$ there are two regimes:
$h+\log(t/2) > f_{out}$ or $< f_{out}$. In
the first regime the determinant is $z$-independent, hence
$G(z)=0$, and we are inside the curves
shown in \ref{f.spectrum}(left),(middle). In the second regime
the determinant is $h$-independent, hence $G(z)$ is also $h$-independent,
and we are outside the curves.

Once we have $f_{out}$ we may calculate the critical parameters
$h_c^{(1,2)}$, for which we have full characterization of the
eigenvalue spectra.
Note that $f_{out}$ depends explicitly on $\VV$. It may happen that
for certain configurations of site randomness $\VV$ the $f$ term wins
while for others the `$h$' term dominates. This leads to a broadening
of the curve of eigenvalues. This effect is difficult to control analytically,
however numerical simulations suggest that this broadening is of order
$1/\sqrt{N}$ \cite{HNUS}. This behavior is reminiscent of the case of
weak nonhermiticity discussed recently in \cite{WEAKLY}.

\subsubsection{Coherent Phase Approximation}

A direct calculation of $G(z)$ is involved. If we were to
make a diagrammatic expansion we would find that a lot of nonplanar
graphs would contribute. Furthermore one cannot do a perturbative
expansion around $h$ small or big, neglecting the `hopping parameters'
above or below the diagonal of the Hamiltonian. This is not surprising,
the HN model is just the nonhermitian version of the Anderson model
in d-dimensions, for which there is no exact solution in large $N$ 
except for the special case of Cauchy randomness (Lloyd model) 
\cite{LLOYD}. 

We have shown in \cite{HNUS} that by assuming the site
randomness and the `hopping' part of the Hamiltonian to be free, we
can just apply the method of  
Blue's functions to the present problem. In this approximation, we resum at
once the planar diagrams and all non-planar diagrams with single-site 
rescattering (some sort of a mean-field analysis). This approximation
bears much in common with the CPA method as we now show. Specifically, let
\eq
H=H_D+H_R
\eqx
where $H_D$ is the deterministic hopping term and 
$H_R$ is the  diagonal random matrix with random entries $V_A$.
Let $G_D$ denote the deterministic Green's function:
\eq
G_D(z)={\rm tr}\, \f{1}{z-H_D} \,.
\eqx
In the CPA the Green's function of the system is given by
\eq
\label{e.sgdef}
G(z)=G_D(z-\Sg)
\eqx
where $\Sg$ satisfies the condition of \cite{TEXTBOOK}
\eq
\cor{ \f{V_A-\Sg}{1-(V_A-\Sg)G(z)} }=0 \,.
\eqx
Multiplying by $G(z)$ and adding and subtracting 1 in the numerator
we get
\eq
\cor{ \f{1}{1-(V_A-\Sg)G(z)} }=1
\eqx
then multiplying by $G(z)$ once again, we obtain
\eq
G(z)=G_R\left( \f{1}{G}+\Sg \right) \,.
\eqx
As a final step let us evaluate the Blue's function for the random
part $B_R$ on both sides of this equation
\eq
B_R(G(z))=\f{1}{G(z)}+\Sg \,.
\eqx
But this is a direct consequence of the addition formula for Blue's
functions
\eq
B(z)=B_R(z)+B_D(z)-\f{1}{z} \,.
\eqx
Namely after reexpressing $B_R$ in terms of $B$ and $B_D$ we obtain
\eq
z-B_D(G(z))+\f{1}{G(z)}=\f{1}{G(z)}+\Sg \,.
\eqx
But now we may use the definition of $\Sg$ (\ref{e.sgdef}) and get
\eq
z-B_D(G_D(z-\Sg))=\Sg
\eqx
which finishes the proof. 

We note that the above arguments do not rely on whether the starting 
Hamiltonian is hermitian or not. In the hermitian case our conclusion
is in agreement with the one of Neu and Speicher \cite{NEUSPEICHER},
who used different arguments. Since the Green's 
functions for the nonhermitian and hermitian ensembles coincide in 
the ``outer'' region,  the addition of the Blue's functions together 
with the balance of the free energies as detailed in the preceding sections,
provide for a concise approximation scheme for nonhermitian localization
in the large $N$ limit. We now
proceed to implement this.

\subsubsection{Nonhermitian Localization}

The Green's function for the deterministic part is just
\eq
G_D(z)=\f{1}{N} \sum_{n=1}^N \f{1}{z-t\cos(2\pi n/N+ih)} =
\left\{ \begin{array}{cl} 1/\sqrt{z^2-t^2} & \mbox{``outside''}\\ 0 &
\mbox{``inside''} \end{array} \right.
\eqx
where the last equality holds in the large $N$ limit.
This shows that we may define the Blue's functions only {\em outside}
the curve of eigenvalues. There it is given by
\eq
B_D(z)=\sqrt{\f{1}{z^2}+t^2} \,.
\eqx
The Blue's function for the site randomness $V_A$ is fixed by the type of
distribution we select. For the (solvable) case of a semicircular
distribution, the Blue's function is 
\eq
B_R^{semicircular}(z)=z+\f{1}{z}
\eqx
while for a uniform distribution of width $\Delta$ it is
\eq
B_R^{uniform}(z)=\Delta \coth \Delta z
\label{uniform}
\eqx
and for a Cauchy distribution\footnote{For ensembles with unbounded
moments, the proof of the addition formulae is more subtle,
cf. \cite{CAUCHYADD}.}
\eq
B_R^{Cauchy}(z)=\f{1}{z}-i\gm \,.
\eqx
Therefore the Blue's function for the HN Hamiltonian is
\eq
B(z)=B_D(z)+B_R(z)-\f{1}{z} \,.
\eqx
So the Green's function (in CPA) would satisfy the
equation $B(G(z))=z$. In particular, for a uniform 
distribution the Green's function satisfies
\be
z=\sqrt{\frac{1}{G^2}+t^2} +\Delta {\rm coth} \Delta G -1/G
\label{Greenuni}
\ee
while for a Cauchy distribution it satisfies
\eq
\sqrt{\f{1}{G^2}+t^2}-i\gm=z \,.
\eqx
Integrating the Blue's function for the HN Hamiltonian gives the free
energy $f_{out}$ (in this case a direct integration of $G$ would be 
infeasible)
\be
f_{out}(z)=\Delta G \coth \Delta G -1 +\log \frac{1+\sqrt{1+G^2}}{2}
-\log \frac{\sinh \Delta G}{\Delta}
\label{freeuni}
\ee
for uniform randomness and 
\eq
f_{out}(z)=\log(i\gm+z+\sqrt{(i\gm+z)^2-t^2})-\log 2
\eqx
for Cauchy randomness.

We will now calculate $h_c^{(1)}$ --- the value of $h$
where the curve of eigenvalues starts to appear (i.e. when point B in
figure \ref{f.spectrum} leaves the origin along the imaginary axis).  
This is determined by the condition that the boundary between the two 
phases passes through $z=0$
\eq
h+\log\f{t}{2}=f_{out}(0) \,.
\label{balance}
\eqx
For a uniform distribution (\ref{freeuni}) the solution of the analytical 
condition (\ref{balance}) is only possible numerically. In figure 
\ref{f.critical} we compare this solution to the {\it exact} numerical 
values for the HN Hamiltonian. The agreement is fairly good. For a Cauchy 
distribution, the critical condition (\ref{balance}) is simply 
\eq
\sinh h_c^{(1)}=\f{\gm}{t} \,.
\eqx
The same condition was obtained by Br\'{e}zin and Zee~\cite{UPDATENH},
and Goldsheid and Khoruzenko \cite{HNOTHER}, using different 
arguments. We note, in agreement with these authors, that
in the case of Cauchy randomness the results are {\it exact}, much like
the Lloyd model in the hermitian case.

\begin{figure}
\centerline{\epsfxsize=12cm \epsfbox{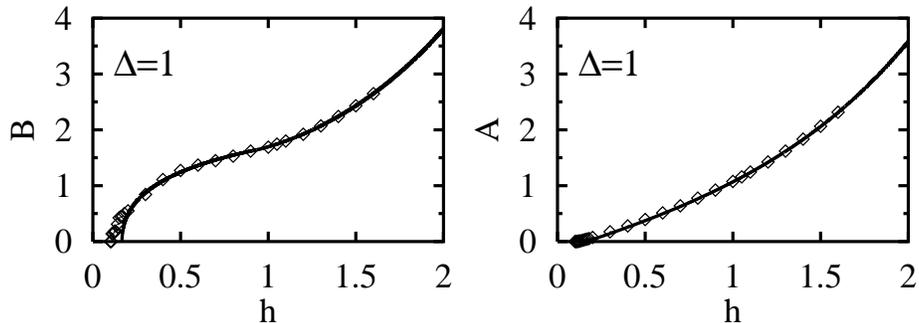}}
\caption{Dependence of A (forking point) and B (top point on the 
imaginary axis upon lift-off at $z=0$) versus $h$ for $\Delta=1$ 
(uniform randomness). The solid lines follow from the analytical 
results using the addition law (CPA).}
\label{f.critical}
\end{figure}

\section{CONCLUSIONS}

Voiculescu's free random variable approach has a simple realization
in terms of Blue's functions, as originally suggested by Zee\cite{ZEE}.
This applies for both the R- and S-transforms. In the first part of these 
lectures, we have gone over the essentials of Blue's function calculus, 
emphasizing the generic aspects which are important in physical settings. 
The addition and multiplication of free random Hamiltonians translates
simply to the addition and multiplication of the pertinent Blue's functions.
More importantly, their extrema are in one-to-one correspondence with the
end-points of pertinent eigenvalue distributions, and their analytical 
structure conditions the singularity structure of appropriate partition 
functions.  Moreover Blue's functions provide a surprising method of
finding the boundary of the support of eigenvalues for nonhermitian
ensembles through conformal mappings.

The free random treatment of matrix models amenable to a $1/N$ 
expansion is usually exact to leading order in $1/N$, hence in
agreement with the treatment based on more conventional methods.
The advantage of the former being its simplicity. For disordered 
models which are not immediately amenable to a $1/N$ expansion,
the free random treatment provides for an approximation that is 
identical to the CPA approximation. The latter has been successfully
applied to the Anderson model \cite{ANDERSON}, and its variants 
\cite{NEUSPEICHER,WEGNER}. 
The free random variable approach goes beyond Gaussian
measures, and applies equally well to random matrix models as well
as quantum models with diagonal disorder \cite{HNUS}.

In the second part of these lectures we have discussed effective QCD
in light of the spontaneous breaking of chiral symmetry in the vacuum. 
Using power counting both in the microscopic and macroscopic limit, we 
have argued that the partition function in the large volume limit is solely 
conditioned by the mode of symmetry breaking on the pertinent coset
span by the little group \cite{GASSER,SMILGA}. The stability of the 
partition function against soft and hard quantum corrections, conditions 
the universal behavior of the spectral oscillation of the QCD Dirac operator 
both at zero virtuality and in the bulk of the infrared region. For chiral
random matrix models this means simply that arbitrary polynomial weights
that are chirally invariant yield by power counting the
same universal spectral oscillations in the vacuum.

The universal behavior of spectral oscillations in effective QCD occurs
in the microscopic limit with light quark masses (for chiral symmetry
to be relevant) and large {\it enough} volumes (for nonperturbative
physics to set in). This is the limit, however, where the vacuum has
no spontaneous alignment yet (the chiral condensate is driven by the
light quark masses). The universality
arguments do not carry any dynamical content. Rather, they
reflect on the kinematical correlations induced by the compactness of the
coset space associated to the would-be Goldstone modes, in the presence
of a universal $(N_f, N_f)$ symmetry breaking pattern.

When these arguments are combined with the universality of the chiral phase 
transition as advanced by Pisarski and Wilczek \cite{PISARSKI} 
in the context of QCD, they show that these oscillations survive the 
effects of matter at the critical point, thereby generalizing the 
arguments in the vacuum. 
Although our arguments were restricted to effective QCD, they apply equally
well to QCD in odd dimensions, as well as effective fermionic theories
with spontaneous symmetry breaking such as the Nambu-Jona-Lasinio models
or the instanton models. For the latter spectra, we note that power counting 
is simplified on the hypertorus. These assertions can be readily checked.

In the third part of these lectures, we have applied some of the 
Blue's function techniques to chiral random matrix models, in the
non-universal regions where the $1/N$ expansion applies. The latters
belong to a cohort of effective models of QCD in the long wavelength limit,
that reproduce the microscopic sum rules. In the macroscopic limit they are 
consistent with power counting only at tree level, hence their limited
range of consistency with the QCD Ward identities. They are however useful 
for addressing issues related to the thermodynamical limit, as well as 
mean-field regions in most of the phase transitions, barring the case 
of large Ginzburg windows \cite{GINZBURG}. In their standard form they are
schematic versions of the instanton model to the QCD vacuum. Unlike
the latter, however, they can be extended to accommodate microscopic QCD
in odd dimensional spaces where the issue of chiral symmetry is subtle,
and the question of instantons obsolete.

As effective models of QCD in the infrared, the chiral
random matrix models  can be used to illustrate
the subtleties between the large volume limit versus the smallness or 
largeness of the QCD external parameters. In this respect, we have analyzed 
the U(1) problem, and the dependence of the vacuum partition function on the
$\theta$ angle. In both cases the large volume limit is subtle when light
quarks are present. Needless to say, that lattice QCD simulations at finite
$\theta$ are not yet available, owing to the complex character of the fermion 
determinant.

We have shown that the concept of addition law of Blue's functions finds
a simple realization when the temperature or quark chemical potential
are schematically inserted in the standard chiral random matrix models.
These models are tied to QCD only in the context of mean-field
universality, and may be given non-standard versions to accommodate for
non-mean-field universalities. As effective models to the QCD phase
transition, they can be regarded as crude variants of the
Nambu-Jona-Lasinio model, or the instanton model, to which they
naturally relate. We stress, however, their simplicity and also their
user-easy access to the quark spectra. Under a phase change, the latters
are expected to undergo structural changes near zero virtuality provided
that the soft and hard modes decouple in the large volume limit. This is
the case for infrared sensitive observables, including order
parameters. In this sense, our comparison to the present lattice data at
the transition point is striking~\cite{USLATTICE}. We note, however,
that the absence of thermal and Fermi motion in these models, makes
their extension to most observables suggestive at best.

Chiral random matrix models of effective QCD are also interesting
because they tie some of the fundamental aspects of QCD to other models
of disorder.  In this context chiral random matrix models at finite
chemical potential resemble closely models of disorder with
dissipation. The analysis in terms of Blue's functions generalize
handily to these cases as well.

We have presented 
models with strong dissipation, currently in use to model quantum chaotic 
scattering with many open channels, as well as models of weak dissipation,
such as the model recently discussed by Hatano and Nelson. The latter is
disordered quantum mechanics in $d$ dimensions, that is still amenable to
random matrix techniques provided that freeness is used. We have shown that 
this approximation is identical to the CPA approximation commonly in use for 
hermitian disordered problems ({\it e.g.} Anderson model).

The models of strong nonhermiticity discussed here, apply to a large
number of physical phenomena. They show up in nuclear, atomic and
molecular spectroscopy, mesoscopic ballistic devices, microwave cavities
and the physics of condensed state chemical reactions, to cite a
few~\cite{SOMMERSREV}.  Our method should shed more light on the already
existing analyses.

The current models of weak nonhermiticity offer new field of investigations 
that are worth pursuing with the present methods. The model of Hatano and 
Nelson~\cite{HATANONELSON} (nonhermitian analogue of the Anderson 
model~\cite{ANDERSON}) for the description  
of flux depinning in type II superconductors offers a good starting point. It
is a canonical example of nonhermitian localization, that may become
relevant to a variety of other models with a directed diffusion,
such as: models for the growth of bacterial populations~\cite{SHNERB},  
models for spreading genetic mutations, models of chemical reactions, 
neural networks, turbulence, and models of growth of monetary capital.
We are confident that Voiculescu's idea of freeness will open 
a vested interest in this rich array of problems.

\section{Acknowledgments}

This work was partially supported by the US DOE grant DE-FG-88ER40388,
by the Polish Government Project (KBN) grants  2P03B19609 and  2P03B04412
and by Hungarian grants OTKA T022931 and FKFP0126/1997.


\end{document}